\documentclass[acmsmall, authorversion, nonacm]{acmart}

\usepackage[utf8]{inputenc}

\usepackage{alltt}
\usepackage{listings}
\usepackage{graphicx}
\usepackage{lscape}
\usepackage{rotating}
\usepackage{pdflscape}
\usepackage{adjustbox}
\usepackage{tabularx}
\usepackage{multirow}
\usepackage{enumitem}

\usepackage{footnote}
\usepackage{footnotebackref}
\usepackage{tablefootnote}
\usepackage{threeparttable}
\usepackage[para]{footmisc}

\usepackage{amsthm}
\newtheorem{definition}{Definition}

\usepackage{diagbox} 

\usepackage[normalem]{ulem}
\newcommand{\change}[2]{\sout{\textcolor{red}{#1}}\textcolor{blue}{#2}}
\renewcommand{\change}[2]{\textcolor{blue!65!black}{#2}}
\renewcommand{\change}[2]{\textcolor{black}{#2}}

\newcolumntype{M}[1]{>{\centering\arraybackslash}m{#1}}

\usepackage[textsize=tiny, textwidth=1.2cm, disable]{todonotes}
\usepackage{hyperref}

\newcommand{\snippet}[1]{``\texttt{#1}''}

\setcopyright{none} 
\copyrightyear{2022}
\acmYear{2022}
\acmDOI{} 

\acmJournal{CSUR}
\acmVolume{0}
\acmNumber{0}
\acmArticle{}
\acmMonth{6}

\makeatletter
\let\@authorsaddresses\@empty
\makeatother

\begin{document}

\fancyfoot[RO,LE]{\footnotesize ACM Computing Surveys. Accepted for publication: April 6th, 2022.}


\title[Survey and Taxonomy of Adversarial Reconnaissance Techniques]{Survey and Taxonomy of\\Adversarial Reconnaissance Techniques}

\author{Shanto Roy}
\email{sroy10@uh.edu}
\affiliation{%
  \institution{University of Houston}
  \city{Houston}
  \state{Texas}
  \postcode{77204}
  \country{USA}
}

\author{Nazia Sharmin}
\affiliation{%
  \institution{University of Texas at El Paso}
  \city{El Paso}
  \state{Texas}
  \postcode{79902}
  \country{USA}
}
\email{nsharmin@miners.utep.edu}

\author{Jaime C. Acosta}
\affiliation{%
  \institution{DEVCOM Army Research Laboratory}
  \city{El Paso}
  \state{Texas}
  \country{USA}
}
\email{jaime.c.acosta.civ@army.mil}

\author{Christopher Kiekintveld}
\affiliation{%
  \institution{University of Texas at El Paso}
  \city{El Paso}
  \state{Texas}
  \postcode{79902}
  \country{USA}
}
\email{cdkiekintveld@miners.utep.edu}

\author{Aron Laszka}
\affiliation{%
 \institution{University of Houston}
 \city{Houston}
  \state{Texas}
  \postcode{77204}
  \country{USA}
}
  \email{alaszka@uh.edu}

\setcounter{page}{1}
\begin{abstract}
Adversaries are often able to penetrate networks and compromise systems by exploiting vulnerabilities in people and systems. The key to the success of these attacks is information that adversaries collect throughout the phases of the cyber kill chain. We summarize and analyze the methods, tactics, and tools that adversaries use to conduct reconnaissance activities throughout the attack process. First, we discuss what types of information adversaries seek, and how and when they can obtain this information. Then, we provide a taxonomy and detailed overview of adversarial reconnaissance techniques. The taxonomy introduces a categorization of reconnaissance techniques based on the source as third-party, human-, and system-based information gathering. This paper provides a comprehensive view of adversarial reconnaissance that can help in understanding and modeling this complex but vital aspect of cyber attacks as well as insights that can improve defensive strategies, such as cyber deception.  
\end{abstract}




\maketitle

\section{Introduction}

Businesses and governments must develop novel capabilities and technologies to stay competitive, but this innovation also leads to new cybersecurity challenges.
As new security measures are developed and implemented, adversaries continuously evolve new tactics and identify new vulnerabilities to conduct attacks. A 2018 Gartner report estimated that security expenses would grow up to $\$124$ billion in total in 2019 \cite{GartnerF27:online}. Another report published by Verizon reveals that 
$33\%$ of data breach involved social engineering, and $28\%$ involved malware \cite{2019data37:online}.
The report also shows that in $56\%$ of the reported breaches, it took months or longer to discover the~attack. Reconnaissance activities 
are 
one of the key stages in conducting successful attacks.

\emph{Reconnaissance} (or \emph{recon}) in cybersecurity refers to the ongoing process used by attackers to gather as much information as possible about target systems or networks that can be used to conduct various types of malicious activity, such as gaining unauthorized access or denial of service. This is a crucial aspect of a successful attack, since the gaps in the security of a well-managed network may be small, and attackers may need to chain together exploits of multiple vulnerabilities to execute highly effective attacks. Better understanding how attackers go about gaining this information can help us to model this key aspect of attacker behavior, and to build better, more targeted defensive strategies for preventing attackers from easily gaining the valuable information that they need for planning attacks.  


\begin{definition}
Reconnaissance is a process (sequence of actions) performed by adversaries to gather information about target networks and systems that is necessary for successfully exploiting vulnerabilities and furthering the adversaries' goals.
\end{definition}

Reconnaissance plays a crucial role throughout the cyber kill chain\footnote{The cyber kill chain describes different stages of a cyber attack. More details are provided in Section~\ref{sec:phase}}.  Adversaries collect information about targets using different tactics, techniques, and procedures (TTP).  Adversaries can gather information sitting outside or inside target networks for months or even years. Sophisticated attackers can utilize multiple reconnaissance techniques while remaining undetected, making it more difficult for defenders to realize when a system is under attack. Several companies and organizations, including FireEye\footnote{\url{https://www.fireeye.com/}},
Cisco\footnote{\url{https://tools.cisco.com/}},
Symantec\footnote{\url{https://www.symantec.com/}},
McAfee\footnote{\url{https://www.mcafee.com/}},
Microsoft\footnote{\url{https://www.microsoft.com/}},
Malwarebytes\footnote{\url{https://www.malwarebytes.com/}},
Bitdefender\footnote{\url{https://www.bitdefender.com/}},
Kaspersky\footnote{\url{https://usa.kaspersky.com/}},
Fortinet\footnote{\url{https://www.fortinet.com/}},
ThreatTrack (Now, Vipre\footnote{\url{https://www.vipre.com/}}),
ISACA\footnote{\url{https://www.isaca.org/}},
and CIS\footnote{\url{https://www.cisecurity.org/}}, 
are involved in investigating and analyzing the TTPs of attackers. However, despite the importance of reconnaissance in understanding attacker behavior, there is relatively little comprehensive academic research published on the reconnaissance process and TTPs. In particular, there is a gap in the literature on surveying, categorizing, and understanding the overall attacker reconnaissance process. We bridge this gap by {collecting and analyzing a broad range of work on adversarial reconnaissance and} building a taxonomy of  reconnaissance activities and techniques that addresses the following main questions:

\begin{itemize}
\item[\textbf{Q1}] \textbf{Reconnaissance Target Information:} What types of information do adversaries seek through reconnaissance? 
\item[\textbf{Q2}] \textbf{Reconnaissance Phases:} {When do adversaries perform reconnaissance?}
\item[\textbf{Q3}] \textbf{Taxonomy of Reconnaissance Techniques:} What are the main categories of reconnaissance techniques {and how do adversaries apply these techniques}? What are the characteristics of these {techniques in terms of what information is obtained and when/how they are utilized?} 
\end{itemize}

We answer the first question with an analysis of the different types of information commonly collected by attackers in the reconnaissance process (Section~\ref{sec:target_info}).
Initially, we categorize the target information in terms of \emph{non-technical} and \emph{technical} information. The non-technical (or social) category includes \emph{organization details} and \emph{people information}. Technical information consists of   \emph{network}, \emph{host machine}, \emph{application}, and \emph{user}-level information. 

{We answer the second question by considering reconnaissance activities in two main parts of the \emph{cyber kill chain.} Here, we {divide} reconnaissance activities into \emph{external} and \emph{internal} reconnaissance.}
External recon is performed from outside the organization's network while internal recon is performed after gaining access to the target network. Internal recon is comparatively more effective in terms of gathering detailed information; however, external recon process has less chance to be identified by the defender.

We answer the {third} question by developing a taxonomy of reconnaissance techniques (Section~\ref{sec:recon_techniques}). We categorize different reconnaissance techniques {and map the techniques with the target data (Section~\ref{sec:target_info}) and phase (Section~\ref{sec:phase})}. We categorize recon techniques {initially based on source: \emph{third-party source}, \emph{human}, and \emph{system}}. {Third-party source-based} target footprinting includes mostly passive techniques that are performed by tracking down the online (Internet) or offline (documents) footprints of targets {that can be obtained from third parties (e.g., public 3rd party websites, or the dark web)}. {Human-based recon techniques, a.k.a.} social engineering, involves active techniques intending to fool people into giving away confidential details or access information. {System-based recon techniques are used to obtain information by observing or interacting with the target system locally (e.g., localhost discovery) or remotely (e.g., network scanning and sniffing).}

\paragraph{Scope}
Reconnaissance is performed not only by adversaries (black/grey hat hackers) but also by security researchers (white hat hackers, blue teams, etc.) for security testing purposes. {We specifically} discuss reconnaissance from the adversary's perspective, broadly focusing on  
targeted attack scenarios {(both large-scale and small-scale attacks)}, including advanced persistent threats. 

We discuss and elaborate the taxonomy  based solely on  reconnaissance procedures; the taxonomy does not cover other steps, techniques, or phases included in a threat model or the cyber kill chain. For example, we do not cover what procedures adversaries follow to compromise a host or to install a malware on a system.
Nonetheless, we cover {different} recon techniques (e.g., social engineering, scanning, {etc.}) that are used to collect technical or non-technical information at both the external and internal recon phases.

\paragraph{Organization}
The rest of this paper is organized as follows: Section~\ref{sec:real_world_cyber_attacks} presents case studies of previous cyber attacks {and how adversarial reconnaissance played an important role in determining attack strategies.} {Section~\ref{sec:related_work} presents} related {surveys and case-studies} regarding {in-general or specific} reconnaissance {techniques and tools}. {Then,} Section~\ref{sec:target_info} provides insight into what information adversaries look for, and {Section~\ref{sec:phase}} discusses when they apply recon techniques during an attack. 
Section~\ref{sec:recon_techniques} categorizes and discusses different  reconnaissance techniques used in both external and internal phases. 
Finally, Section~\ref{sec:concl} 
concludes the paper by summarizing our findings and highlighting research gaps and opportunities in modeling reconnaissance  and developing countermeasures. 

\section{Case Studies of Real-World Cyberattacks}
\label{sec:real_world_cyber_attacks}

Reconnaissance enables attackers to understand system configurations and to find alternative ways to exploit a system. To illustrate this, we present an example of an advanced persistent threat, called \emph{APT41}, which has been responsible for several cyberattacks since 2014. Then, we discuss two additional well-documented cases that caused tremendous losses as examples of how reconnaissance is important to launching a successful attack {and the types of methods and information involved}. 


\subsection{APT41: Advanced Persistent Threats Analysis}
An \emph{advanced persistent threat} (APT) is a stealthy computer network threat actor that uses clandestine, evasive, continuous, and sophisticated cyberattacks to gain and maintain unauthorized access to a system for a prolonged period without getting detected~\cite{daly2009advanced}.
Usually, the purpose of an APT is to steal sensitive information by monitoring, intercepting, and relaying it rather than causing network outage, denial of service, or infecting systems with malware. 
What differentiates APTs from other attacks are the TTPs that they employ and how the illegally obtained information is used to satisfy the ulterior motives of the threat actors. 
For example, APT41~\cite{ReportDo23:online}, a Chinese espionage operator, targets healthcare, technology, telecommunications, travel services, news, and media firms.
These sectors play crucial roles in 
China's five-year economic development plan~\cite{The13thF90:online}. 
The group injects malicious code into files then signs them with stolen legitimate code-signing certificates. This kind of attack affects a large number of hosts across the world after the distribution of the package. {Using a technique called Execution Guardrail,} information collected from a host (OS version, IP address, Active Directory name, shared network name, etc.) can be used to limit the activation of malware. Once a host has been compromised, the group used techniques like automated collection, data from information repositories, data from local systems, input captures, and screen captures to collect surveillance data that serves their interests.

\subsection{Cyberattack on the Ukrainian Power Grid}
On December 23, 2015, a power grid in Ukraine was compromised by a cyberattack causing a service outage to the customers. The duration of the outage  was only 6 hours. However, it took months to recover from the attack as most of the device firmware was overwritten with {malware}. 
The reconnaissance phase started much earlier, and the group of attackers initially utilized spear-phishing (water-hole attack) and email spoofing attacks to send emails to company workers with a malicious document attached. When a user opened the document, a pop-up menu appeared asking if the user would like to enable a macro; if the user agreed the macro installed a backdoor. The attackers potentially gained  user credentials to log in to the system remotely. As there was no two-factor authentication, it was easy for the attackers to gain access as regular workers. Thereafter, they studied the whole network using
\emph{internal reconnaissance} process for six consecutive months before launching the attack on December, 2015~\cite{case2016analysis}.
The reconnaissance included mostly network and system scanning~\cite{BlackEne6:online, Blackenergy:online}, and discovered field devices including serial-to-Ethernet devices that helped to interpret commands from the deployed SCADA network to the substation control~systems. 

\subsection{Cyber Heist at Bangladesh Bank }
There are also threat groups who are financially motivated to perform cyber heists. For example, the Bangladesh Bank cyber heist caused $\$81$ million in losses \cite{hill2018swift}. \emph{APT38} was the threat actor behind this heist~\cite{ReportAP68:online}, which was well planned to mitigate risks. It has performed some of the biggest cyber heists in the history of cyber crime~\cite{NorthKor13:online}. Before an attack campaign, the group conducted an extensive level of reconnaissance on the target system's personnel for watering hole attacks~\cite{alrwais2016catching}.
In one instance, the group targeted a manager's mailbox to learn about employees who have access to SWIFT servers. SWIFT (Society for Worldwide Interbank Financial Telecommunication) enables secure transactions among financial institutions. The group performed reconnaissance on a bank's remote connection to a third-party vendor with access to the SWIFT servers, which the group later utilized to build their malware.
The group also performed prolonged reconnaissance of network activity and collected user and system information. In one case, it sent LinkedIn invitations to employees who were later targeted in watering hole attacks. 
Once the group had a foothold inside a network, it spent a prolonged period performing reconnaissance over the network---in some cases for two years---before starting fraudulent SWIFT transactions. The group exploited persistent access for as long as it took to learn network topology, permissions, monitoring software, and SWIFT systems. 
They also took control of \emph{sysmon} and \emph{sysinternal} utilities for internal monitoring. 

\paragraph{Lessons Learned}
By analyzing real-world attack scenarios, we see that both external {(spear-phishing or water-hole attacks in the Ukrainian power-grid cyberattack and the Bangladesh Bank cyber heist)} and internal reconnaissance {(Execution Guardrail techniques of APT41, system scanning in the Ukrainian power-grid cyberattack)} play a significant role in a successful attack. The threat groups initially collect publicly-available information, extract necessary details, and plan accordingly. Then, they gain access by breaching internal systems and use malware to gain access in the internal network, followed by obtaining system details. Advanced persistent threats can perform internal reconnaissance for a long time {(six months during the Ukrainian power-grid cyberattack)} without being detected; and in the meantime, the adversaries keep finding loopholes to improve their attack plan. Based on {these case studies}, we can see that many attacks are well-planned and cause tremendous loss to the target organizations.

\section{Related Literature Surveys}
\label{sec:related_work}

We now discuss previous survey papers that discuss
different aspects of cyber reconnaissance in terms of techniques and tools.
The number of reviews {that focus specifically} on reconnaissance is relatively low. Some studies have surveyed and discussed different reconnaissance techniques, methodologies, and approaches (e.g.,~\cite{mazurczyk2021cyber,rai2021using,hassan2018evolution,shaikh2008network,shaikh2008network,bhuyan2011surveying,dar2018silent}).
Other works have evaluated the performance of publicly available reconnaissance tools (e.g.,~\cite{holm2011quantitative,wang2017ethical,coffey2018vulnerability}).

\subsection{Surveys of Reconnaissance Techniques}

{A few previous papers~\cite{mazurczyk2021cyber,dar2018silent,sanghvi2013cyber} have attempted to present adversarial reconnaissance techniques comprehensively. For example, Mazurczyk et al.\ classified the evolution of cyber reconnaissance into four categories: internet intelligence, network information gathering, side-channel attacks, and social engineering~\cite{mazurczyk2021cyber}. They also provide examples of specific techniques based on the level of interaction and evolution over time (older vs. newer techniques). However, the categorization is not comprehensive (not all types of recon techniques are mentioned, e.g., sniffing and localhost discoveries) and did not provide a clear and concise taxonomy. The authors also discussed human-based countermeasures (awareness), reactive countermeasures (sniffing and side-channel prevention), and proactive countermeasures (cyber deception and moving target defense) to mitigate reconnaissance.  The two other papers focused on network-based reconnaissance techniques and did not include other reconnaissance techniques. Next, we discuss survey papers that focus on and categorize specific types of reconnaissance techniques.}

\paragraph{Open Source Intelligence}
{There are few works~\cite{glassman2012intelligence,kanta2020survey,tabatabaei2016osint,hassan2018evolution,rai2021using} that survey different techniques in \emph{open source intelligence} (OSINT) from the perspective of cyber security. Glassman et al. discussed how the world wide web provides access to immense information that can be potentially used for decision making and problem solving~\cite{glassman2012intelligence}. Tabatabaei et al. listed several tools that can be used for the collection, storage, and classification of open-source data~\cite{tabatabaei2016osint} in the context of security. Some other papers discussed OSINT for a specific purpose such as reliable web searching~\cite{rai2021using} or password cracking~\cite{kanta2020survey}}.

\paragraph{Social Engineering}
{Several papers have presented taxonomies of different \emph{social engineering} (SE) attacks~\cite{alharthi2020taxonomy,chiew2018survey,heartfield2015taxonomy,salahdine2019social,gupta2016literature}. Alharthi et al. categorized the techniques in two types: technical, where the attacker uses media (e.g., mobile text, phishing site) to manipulate the user to reveal sensitive data, and non-technical, where the attacker directly interacts with the target. Salahdine et al. also classified the SE attacks as \emph{human-based} and \emph{computer-based}~\cite{salahdine2019social}. Heartfield et al. presented a taxonomy of semantic attack mechanisms of different social engineering attacks where they defined attack characteristics in three stages: orchestration, exploitation, and execution~\cite{heartfield2015taxonomy}. Other works surveyed specific SE techniques such as phishing~\cite{chiew2018survey,gupta2016literature}.}

\paragraph{Cyber Scanning}
{There are a number of works that discussed cyber scanning techniques~\cite{shaikh2008network,bhuyan2011surveying,bou2013cyber,claypool2002stealth,barnett2008towards,de1999review}.}
Shaikh et al.\ provided a general overview of reconnaissance techniques that focuses on the classification of probes and discusses methods of reconnaissance including surveillance, eavesdropping, and intercepting communications~\cite{shaikh2008network}. The authors classified probes into three groups: host detection, port enumeration, and vulnerability assessment. They also highlighted several detection challenges and approaches for counter-probing activities.
Some studies have focused on scanning techniques, such as Arkin's work, which reviewed scanning techniques concentrating on ping sweeps, port scans, and operating system identification~\cite{arkin1999network}. Meanwhile, {Claypool conducted a study on stealthy port scanning methods~\cite{claypool2002stealth}. He discussed half-open scan, Xmas tree scan, UDP scan, Null scan, Fragmentation, Decoying, and Spoofing, which are popular forms of stealthy scanning techniques.} Vivo et al.\ reviewed TCP
port scanners, several scanning techniques developed to bypass firewalls analysis and filtering, stealth scanning, basics
of UDP
scanning, and scanning related to specific application-level protocols~\cite{de1999review}. Bhuyan et al.\ surveyed and discussed the effects of frequent port scan attacks~\cite{bhuyan2011surveying}. A comparison of port scan methods based on type, mode of detection, mechanisms used for detection, and other characteristics were discussed in detail. 

\paragraph{Side-channel Attacks}
{Surveys of side-channel attacks~\cite{spreitzer2017systematic,sayakkara2019survey,lyu2018survey,hettwer2020applications} have categorized these techniques based on different dimensions: active vs.\ passive, logical properties vs.\ physical properties, and local vs.\ vicinity vs.\ remote~\cite{spreitzer2017systematic}. All of these categorizations have overlapping techniques and there is no clearly preferable way to categorize different side-channel attacks. Some works surveyed different side-channel attacks for specific techniques such as cache~\cite{lyu2018survey} and electromagnetic emission~\cite{sayakkara2019survey} or applications of different machine learning techniques in side-channel attacks~\cite{hettwer2020applications}.}

\paragraph{Summary}

{We have described several survey papers that focus on specific types of reconnaissance techniques (e.g., open-source intelligence, social engineering, scanning, sniffing, and side-channel attacks). However, to the best of our knowledge, no comprehensive survey work~\cite{mazurczyk2021cyber,dar2018silent,sanghvi2013cyber} categorizes and presents a taxonomy of all types of reconnaissance techniques, with clear categorization and distinction. Table~\ref{tab:lit_comparison} presents a comparison between our work and other general reconnaissance surveys in terms of which techniques are discussed and which are not. {Our main objective in this paper is to comprehensively describe} what target data adversaries are looking for, when and how they perform recon, {and to provide a clear} taxonomy of recon techniques.}

\begin{table}[!ht]
\centering
\caption{Comparison of Existing Reconnaissance Surveys}
\label{tab:lit_comparison}
\resizebox{\textwidth}{!}{

\begin{tabular}{|c|c|c|c|c|c|c|}
\hline
\diagbox[width=8em]{Works}{Techniques} & \begin{tabular}[c]{@{}c@{}}Open Source\\Intelligence\end{tabular} &  \begin{tabular}[c]{@{}c@{}}Social\\Engineering\end{tabular} & \begin{tabular}[c]{@{}c@{}}Cyber\\Scanning\end{tabular} & \begin{tabular}[c]{@{}c@{}}Sniffing\end{tabular} & \begin{tabular}[c]{@{}c@{}}Host\\Discovery\end{tabular} & \begin{tabular}[c]{@{}c@{}}Side-Channel\\Attacks\end{tabular} \\ \hline\hline
 
\begin{tabular}[c]{@{}c@{}} Sangvi et al., 2013\end{tabular} & \text{\sffamily X}  & \text{\sffamily X} & \checkmark & \text{\sffamily X} & \text{\sffamily X} & \text{\sffamily X} \\ \hline

\begin{tabular}[c]{@{}c@{}} Dar et al., 2018\end{tabular} & \text{\sffamily X}  & \text{\sffamily X} & \checkmark & \text{\sffamily X} & \text{\sffamily X} & \text{\sffamily X} \\ \hline

\begin{tabular}[c]{@{}c@{}} Mazurczyk et al., 2021\end{tabular}  & \checkmark & \checkmark & \checkmark & \text{\sffamily X} & \text{\sffamily X} & \checkmark\\ \hline

\begin{tabular}[c]{@{}c@{}} Our Work \end{tabular} & \checkmark  & \checkmark & \checkmark & \checkmark & \checkmark & \checkmark\\ \hline

\end{tabular}
}
\end{table}

\subsection{Surveys of Reconnaissance Tools}

Tundis et al.\ discussed varieties of vulnerability analysis tools~\cite{tundis2018review} and provided corresponding qualitative analysis including the advantages and disadvantages of these tools. 
The work was not limited to finding available tools and procedures to recon a typical system, but it also provided a comprehensive overview of how adversaries collect information about  various networks.
Wang et al.\ reviewed the state of the art of  open-source vulnerability scanning tools~\cite{wang2017ethical}. The authors built a virtual lab environment and analyzed the virtual network with Nmap, Nessus\footnote{\url{https://www.tenable.com/products/nessus}}, Retina CS Community\footnote{\url{https://sourceforge.net/projects/retinacommunity/}}, OpenVAS\footnote{\url{https://www.openvas.org/}}, Microsoft Baseline Security Analyzer (MBSA), and Nexpose Community Edition\footnote{\url{https://www.rapid7.com/products/nexpose/}}. Dar et al. investigated several tools, e.g., DNSEnum\footnote{\url{https://github.com/fwaeytens/dnsenum}}, NMap\footnote{\url{https://nmap.org/}}, ZENMap\footnote{\url{https://nmap.org/zenmap/}}, DNSstuff\footnote{\url{https://www.dnsstuff.com/}}, MxToolbox\footnote{\url{https://mxtoolbox.com/}} and experimented on a variety of operating systems. They concluded that DNSEnum, NMap, ZENMap performed well in active reconnaissance and DNSstuff, MxToolbox in passive reconnaissance if adversaries want to keep their identity hidden.

Holm et al.\ analyzed the performance of seven popular scanners AVDS\footnote{\url{https://beyondsecurity.com/avds.html}}, McAfee, Nessus, NexPose, Patchlink\footnote{\url{https://www.ivanti.com/solutions/security}}, QualysGuard\footnote{\url{https://www.qualys.com/qualysguard/}}, and SAINT\footnote{\url{https://www.carson-saint.com/}}~\cite{holm2011quantitative} on a network consisting of 20 physical servers running a total of 28 virtual machines with various operating systems and versions.  
In another work, Coffey et al.\ analyzed various network scanning tools against SCADA equipment to
examine the differences between issue identification and asset discovery~\cite{coffey2018vulnerability}. 
The authors experimented on {ICS} and {SCADA} systems by finding vulnerabilities using the same scanning tools that are employed on conventional IP networks, and suggested developing a network scanner that is capable of obtaining information from both serial and Ethernet devices at the same~time.

\paragraph{Summary}
We find several research works that survey and discuss different aspects of reconnaissance. Some of these categorize techniques while others evaluate related tools in terms of effectiveness. However, none of these existing works provide a comprehensive overview of the entire reconnaissance process; rather, they each focus on certain parts of the process. Therefore, in this paper, we address this gap by providing a comprehensive survey and taxonomy of reconnaissance techniques, which considers information gathering procedures throughout the entire cyber~attack process.
%


\section{Reconnaissance Target Information}\label{sec:target_info}

\begin{figure}[!ht]
    \centering
    \includegraphics[width=.8\textwidth]{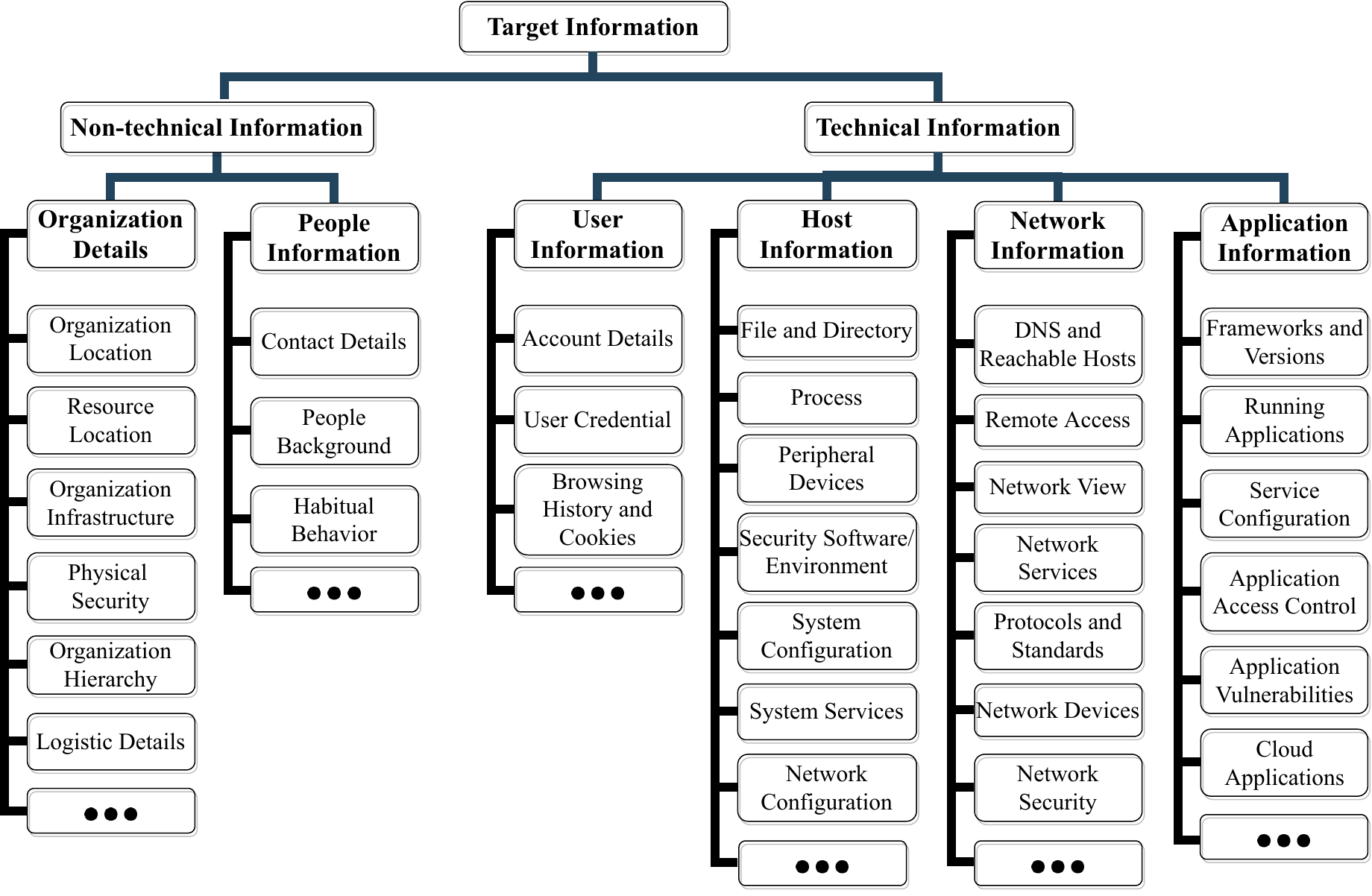}
    \caption{Categories of Target Information for Reconnaissance}
    \label{fig:recon_taxonomy_info}
\end{figure}

Adversaries look for different types of information throughout the attack process. Target information is highly interconnected, and adversaries may need to acquire it sequentially. 
{Furthermore, the type of information the attacker needs} also depends largely on the adversary's objectives {and capabilities}. We categorize the information that adversaries look for during a large-scale network breach. We consider primarily large-scale attacks since they cover an expansive scope of the information that may be acquired by  sophisticated adversaries.
Figure~\ref{fig:recon_taxonomy_info} presents our categorization of the types of information that adversaries look for while performing an attack. 
We divide the adversary's target information in two main {types}: \emph{Non-technical (Social) Information} and \emph{Technical Information}. {The reasoning for this high-level categorization is that adversaries use these types of information for different types of attacks. Non-technical information (e.g., people contact details, physical security, etc.) is often most useful for performing social engineering and initial access planning. Technical information (e.g., host or network configurations) is helpful for adversaries to find vulnerabilities to compromise specific systems, escalate privileges, establish durable footholds, move laterally in networks, and achieve specific objectives.}

\subsection{Non-Technical Information} 
Non-technical or 
social information includes details about the target organization, it's physical
infrastructure, business processes, logistic details, and most importantly, potential vulnerabilities (e.g., flaws in physical security systems or building access control).
Information regarding people who are employees or members of the target organization is another crucial element since adversaries can use this information to trick people into giving away confidential information or granting access to resources~\cite{krombholz2015advanced}.

\subsubsection{Organization Background and Details}
Organization information includes the organization's background, resources, employee contacts and work details, 
physical access and security policies, etc.~\cite{tabatabaei2016osint}. 
Whether 
adversaries target a particular organization depends primarily on the organization's resources 
and if those resources are valuable, vulnerable, and accessible at the same time. 
The security of technical assets depends in part on physical security mechanisms since gaining physical access can be a viable approach for compromising security~\cite{phua2009protecting}.
Publicly available resources are one of the primary {initial} data sources for adversaries.  

\begin{itemize}[topsep=2pt, leftmargin=1em]
    \item \textbf{Physical Attributes:}
    Adversaries can attempt to discover physical attributes of an organization such as location, physical infrastructure, physical security systems, physical resource locations and organization, and resource accessibility~\cite{reconkil90:online}. Adequate information can lead to effective social engineering attacks, such as gaining physical access using reverse social engineering~\cite{case2016analysis}.

    \item \textbf{Logistics Details:} Adversaries can look for logistics information such as financial and business processes or intelligence, employee and management hierarchy, resource arrangement, and other activities~\cite{tabatabaei2016osint}. Supply chain management is also important since it may leak important data regarding the organization~\cite{boyson2014cyber}.
    Adversaries have also been reported to inspect and steal data from the third parties~\cite{Chineseh20:online}.

\end{itemize}

Much of the organizational information is available online (e.g., on business websites. News and blogs are also considered reliable sources for providing an outline and profile of an organization~\cite{tabatabaei2016osint}. With increasing communication through social media, obtaining organizational information has become easier. Adversaries can also join the organization and access confidential information as an insider~\cite{nykodym2005criminal}.

\subsubsection{Personal Information}
Personal information about people such as contact details, technical or financial background, habits, and behavioral traits, are information that adversaries attempt to collect in order to analyze people's weaknesses. Finding these weaknesses is useful for applying social engineering techniques to gain remote access to the victims' machines or online~accounts~\cite{albladi2018user}.

\begin{itemize}[topsep=2pt, leftmargin=1em]
    \item \textbf{Contact Details:} 
    Adversaries can collect contact details such as email addresses, phone numbers, identity information, etc.\ For example, the  theHarvester\footnote{\url{https://github.com/laramies/theHarvester}} is an open-source tool that
    can collect email addresses given a domain name. A user's contact addresses may also be found in social media and personal or organization websites~\cite{krombholz2015advanced, ivaturi2011taxonomy}. 
    
    \item \textbf{Personal Background:} 
    Information related to the technical or financial background of  people is also useful to adversaries for crafting social engineering attacks~\cite{krombholz2015advanced, ivaturi2011taxonomy}. The technical background of a person reveals what information and organization resources they may have access to.  Technical background can be found on the organization website, an employee's LinkedIn\footnote{\url{https://www.linkedin.com}} profile, BeenVerified\footnote{\url{https://www.beenverified.com/}} report, or a curriculum vitae. 
    
    \item \textbf{Habitual Behavior:}
    Adversaries can also attempt to learn their targets' habits in order to perform social engineering (e.g., phishing) attacks~\cite{albladi2018user}. Sophisticated adversaries can even track habitual social media usage to deceive people, for example based on Facebook usage~\cite{vishwanath2014habitual}.
    
    \item \textbf{Emotional States and Blackmail:}
    Adversaries can also try to observe people's emotional states. 
 For instance, adversaries have been reported to spy on people through webcams in compromised hosts~\cite{brocker2014iseeyou}.
Further, adversaries may take photos of victims to blackmail them
~\cite{lightfoot2016law}.
    
\end{itemize}

\subsection{Technical Information}
Technical details include diverse information about  
networks, hosts, applications, and users. Technical details are especially useful once adversaries have access to the target organization's internal network. Basic technical information can be obtained from an external network, but adversaries  usually need to breach the target network or system to be able to gather more accurate details.

\subsubsection{Network-Level Information}
Adversaries look for network-level information, such as the network topology, network protocols, devices, and services, to understand the local network~\cite{barnett2008towards}. Scanning and sniffing are highly effective approaches for obtaining target information at the network level. Adversaries can also look for network security measures, such as the presence of firewalls or intrusion detection systems. Here, we discuss the most common network-level information that adversaries attempt to obtain.
\begin{itemize}[topsep=2pt, leftmargin=1em]
    \item \textbf{Domain Names:}
    Domain and hostnames are identifiers that adversaries can use to tell which  hosts belong to a particular domain. For example, hosts within the domain \snippet{example.com} may have hostnames
    \snippet{host1.example.com}, \snippet{host2.example.com}, etc. Adversaries can use domain names associated with a particular organization to find extensive 
    technical (e.g., subdomains, standard records such as SOA, MX, etc.) and personal details (e.g., admin contacts)~\cite{hassan2018technical}. 
    
    \item \textbf{Remote Hosts and Network Topology:} 
    Adversaries may try to obtain the reachable IP addresses of either the external (public-facing) or the internal network. Reachable IP addresses can be identified through Internet Control Message Protocol (ICMP) or communication protocols such as TCP or UDP (e.g., APT: 
    OSinfo \cite{Buckeyec50:online}). 
     Sometimes, the list of reachable IP addresses helps adversaries to map the whole network view {(hosts, routers, switches, firewalls, and other network devices)}. 

    
    \item \textbf{Network Protocols and Services:}
    A server can run a wide range of network protocols and services. For example, a server may provide web service (e.g., HTTP),
     file transfer service (e.g., FTP),
    name service (e.g., DNS), 
    mail service (e.g., SMTP),
    etc. For public-facing servers, running services can be identified from outside the organization's network by interacting with the server or sniffing packets. The same objective is achievable for internal servers if adversaries have access to compromised hosts on the internal network (e.g., APT: FIN6 \cite{rptfin6p3:online}).

    \item \textbf{Network Devices:} 
    Adversaries can look for network device information such as  hardware device manufacturer or vendor, operating systems and version, manufacturer settings, networking configurations, etc.~\cite{shaikh2008network, shamsi2014hershel, owens2011non, bifulco2015fingerprinting}. Device information is useful for adversaries when employing exploits that target known vulnerabilities. Numerous tools are available (e.g., Network Watcher\footnote{\url{http://www.nirsoft.net/utils/wireless_network_watcher.html}})
    for  identifying device information, such as the manufacturing company.

    \item \textbf{Network Security:} Organizations often implement  network security measures including firewalls, intrusion detection systems, network zone isolation, honeypots, etc.\ to prevent, detect, and mitigate attacks. Firewall rules define the filtering of inbound and outbound packets for a node or network; zone isolation is a layer-wise security measure; honeypots can identify the presence of intruders by analyzing network traffic or resource request behavior. Adversaries can avoid signature-based detection {(recognizing the signatures of known malware)} using zero-day exploits and try to avoid anomaly detection {(detecting a deviation from normal system or network behavior)} using
    stealthier techniques, such as slower scanning for reconnaissance~\cite{claypool2002stealth}.
%
\end{itemize}

Network-level information is essential for planning remote attacks to penetrate an organization's network, and for lateral movement and avoiding detection once an internal network is breached. 
Modern botnet-based attacks typically compromise systems
remotely and then maintain command channels to execute commands on the compromised systems~\cite{jacob2011jackstraws}.
Channels include various protocols (e.g., Telnet, SSH) used by the remote shell client software. Adversaries can obtain network information using  
network or Internet footprinting (Section~\ref{subsec:footprinting}),
scanning or fingerprinting techniques (Section~\ref{subsec:scanning}), and social engineering (Section~\ref{subsec:social_engineering}).


\subsubsection{Host-Level Information}
Host-level information (such as software configurations, running processes, files and directories, and security environments) is very useful to adversaries for performing 
the next stages of attacks. Specific details can be obtained once a host machine is compromised. Here, we list the most common host-level information that adversaries look for.

\begin{itemize}[topsep=2pt, leftmargin=1em]

    \item \textbf{System Processes:} Information regarding details of installed software, presence of security software or environments, development frameworks, resource location, hardware and software configurations, application setup environment, etc.\ (e.g., APT: 
    APT1 \cite{APT1Expo92:online}, 
    OilRig \cite{TheOilRi58:online}) 
    can be obtained by monitoring and enumerating running processes.
    Process discovery on a compromised machine reveals the list of running processes and services of the system (e.g., APT:
    GravityRAT~\cite{TalosBlo98:online}). 

    \item \textbf{System Platform:} {The type of operating system and its version are crucial factors in security;} using old versions creates more opportunities for attackers to utilize known tools to exploit. Apart from  version identification, adversaries are able to collect OS build type, serial number and installation date (e.g., APT: PowerDuke \cite{PowerDuk79:online}),
    BIOS information (e.g., APT: BlackEnergy \cite{BE2Custo43:online}).
    

    \item \textbf{System Configuration:}
    System configuration includes a wide range of settings from system services to hardware settings. 
    Adversaries can gather information from the Windows registry system using remote access tools and can learn about running programs, their configurations, presence of antivirus or sandbox, etc. 
    {Adversaries can collect hardware information such as CPU speed from a particular registry value (e.g., Trojan: Trojan.Hydraq~\cite{TrojanHy20:online}) and system manufacturer's value from the registry to identify the type of the machine (e.g., Group: Group 123~\cite{TalosBlo98:online}) as well.}
    
    
    \item \textbf{System Hardware and Peripheral Devices:} Hardware details including CPU, primary memory, secondary storage, network card,  video card, {and peripheral devices (e.g., USB or Bluetooth devices)} etc.\ may constitute useful information for  learning about the vendors, virtual machines, and forensic setups. Device information helps adversaries to identify known vulnerabilities in a vendor's product and thus to devise exploitation strategies.
    Adversaries can collect more information about particular hardware such as processor (e.g., APT: FALLCHILL \cite{HIDDENCO0:online}), processor architecture (e.g., APT: DarkHotel \cite{Darkhote57:online}),
    motherboard (e.g., APT: BlackEnergy \cite{BE2Custo43:online}, OopsIE \cite{OilRigta41:online}), 
    primary memory (e.g., APT: DarkComet \cite{DARKCOME17:online}) and drive/volume information (e.g., APT: 
    RunningRAT \cite{GoldDrag14:online}), 
    video cards (e.g., APT: Agent Tesla \cite{InDepthA86:online}),
    and peripheral devices like keyboards (e.g., APT: 
    SynAck \cite{SynAckta1:online}). 
    

    \item \textbf{Security Environment:} 
    Adversaries can learn about {security} environments {(e.g., virtualization or sandbox)} by querying registry values, system services, BIOS information, process list, and system information, such as hardware configuration. Usually, malware is executed after sandbox/VM evasion techniques. Security information includes firewall rules, presence of antivirus, honeypot or sandbox setup, virtualization environment, etc.\ (e.g., APT: DarkHotel~\cite{Darkhote57:online}).
    
    
    \item \textbf{Files and Directories:} Adversaries can look for directory contents and file lists (e.g., APT: Brave Prince \cite{GoldDrag14:online}).
    Particular directories containing configuration information or files with specific extensions (e.g., APT: Microspia \cite{Micropsi75:online})
    can be useful for extracting information about user accounts, password management, software or application configurations, network configurations, etc. Adversaries can also look for users' personal files, financial reports, and proprietary data. 
    
    
\end{itemize}

\subsubsection{Application-Level Information}
Security vulnerabilities at the application level depend largely on three factors: {exploitability}, {detectability}, and {impact of damage}. 
To exploit {application-level} vulnerabilities {(e.g., SQL injection, cross-site scripting, broken access control, etc.)}, adversaries collect application-level information from a system or a network. Here, we list the most common application-level information adversaries look for.

\begin{itemize}[topsep=2pt, leftmargin=1em]

    \item \textbf{Frameworks and Environments:} 
    Hosts run various development frameworks (e.g. web-based frameworks such as Laravel\footnote{\url{https://laravel.com/}} or Django\footnote{\url{https://www.django-cms.org/en/}}) and environments (e.g., application runtime environments such as Java VM), which may have vulnerabilities.     
    Misconfiguration is another possible weakness that creates loopholes and attract adversaries~\cite{eshete2011early}. 
    Therefore, adversaries may attempt to collect the names, version, and runtime configuration information of frameworks that are installed on a system. 
    
    \item \textbf{Security Tools and Applications:}
    Presence of anti-malware and forensic tools may be identified by querying the default software installation directory (e.g., ``Program Files'' on a Windows system 
    (e.g., APT: Astaroth~\cite{WereSee72:online}))
    or by querying registry (e.g., APT: FIN8~\cite{SpearPhi48:online}),
    and running processes (e.g., APT: Darkhotel~\cite{Darkhote57:online}). 
    
    \item \textbf{Application or Package Configuration:} 
    Adversaries may also be interested in learning about the configuration of installed software and applications on a host~\cite{baset2017usable}.  
    Depending on the obtained information {(e.g., versions)}, adversaries may utilize a database of existing exploits available on the dark web or develop exploits themselves. Application configuration information can also reveal access tokens {and} user credentials.
    
    \item \textbf{Cloud Dashboard and API:}
    Adversaries can gather information about virtual machines, cloud tools, services, and other cloud assets that are accessible from the compromised host~\cite{baset2017usable}. Information related to Amazon AWS, Google Cloud Platform, Microsoft Azure, and other popular cloud service configurations can be queried or accessed using dashboards API and 
    command-line interfaces\footnote{\url{https://github.com/RhinoSecurityLabs/pacu}}\textsuperscript{,}\footnote{\url{https://cloud.google.com/security-command-center/docs/quickstart-scc-dashboard}}.
    
    \item \textbf{Databases:}
    Database systems are prone to have misconfiguration and human errors that leave systems vulnerable to attacks~\cite{dietrich2018investigating}.
    Adversaries can fingerprint versions
    of MySQL, PostgreSQL, Microsoft SQL Server, and Oracle Database by performing advanced queries~\cite{bertino2007profiling}.
    Advanced remote attackers can also identify the state of an application database, e.g., they can check if the target machine's antivirus signature is updated~\cite{al2011application}.
    
    \item \textbf{GUIs:}
    Apart from this information, adversaries can also obtain data from the GUI windows of running applications. For example, they can collect window titles (e.g., APT: Remexi~\cite{Chaferus79:online}) or text content (e.g., APT: PowerDuke~\cite{PowerDuk79:online}). Adversaries are also capable of enumerating application windows (e.g., APT: SOUNDBITE~\cite{CyberEsp50:online}) and capturing screenshots of them (e.g., APT: Catchamas~\cite{Infostea64:online}).
    
\end{itemize}

\subsubsection{User-Level Information}
User-level information such as 
account details and access credentials are useful for {everything from}
gaining initial foothold in an internal network to privilege escalation on a compromised host. 
Often, adversaries collect information about user accounts and then try brute-force or dictionary-based attacks to gain access~\cite{narayanan2005fast}.

\begin{itemize}[topsep=2pt, leftmargin=1em]
    
    
    \item \textbf{Account Details:} 
User and group information includes the list of users and groups, their login types, access control policies, group permissions, etc. 
APTs can gather information about domain and account information (e.g., account ID, token information, etc.) by observing the list of running processes~\cite{HiddenCo22:online}.
Some APTs are also capable of querying information from account
associated directories and enumerating local and domain users~\cite{Buckeyec50:online}.
    
    
    \item \textbf{User Credentials:}   Some of the most common practices of obtaining user credentials are performing social engineering attacks (e.g., phishing) against target users and installing keyloggers on the users' machines~\cite{bhardwaj2020keyloggers} or utilizing spyware to collect user profile data or login information stored in a browser cache~\cite{ICITBrie94:online} (e.g., APT: Machete~\cite{ESETMach82:online}).
    Adversaries can also take advantage of web browser vulnerabilities to collect user-level information, e.g., by installing a malware extension~\cite{ter2008enhancing} and stealing sensitive information when the user fills out a web form {or from a browser cache~\cite{zhao2013all}}. 
    This has the potential for compromising other services since users often use the same passwords for multiple accounts. 
\end{itemize}

\section{Reconnaissance Phases}\label{sec:phase}

Reconnaissance is present in different forms throughout {the attack} process, and provides key information that is needed to execute subsequent phases.  
Typically, an adversary first selects the target organization and then collects as much information as possible regarding the technical and non-technical features of the target organization using externally available sources to create an effective plan for initial access~\cite{hassan2018technical}. Once adversaries have access to the internal network, they seek more information about the network to engage in lateral movement and compromise other resources~\cite{yadav2015technical}. 
Sophisticated APTs are capable of staying inside their target networks for extended periods of time~\cite{case2016analysis}. 
Ongoing lateral movement with internal discoveries results in continuously expanding access and capability to affect the~target.  

{To understand the adversary's TTP, Lockheed Martin has developed a model called the Cyber Kill Chain\footnote{\url{https://www.lockheedmartin.com/en-us/capabilities/cyber/cyber-kill-chain.html}}, 
which describes the technical aspects and a sequential step-by-step model to understand the movement of APTs \cite{yadav2015technical}. 
However, the basic model does not provide much detailed insight about the reconnaissance (e.g., internal scanning and discovery) processes throughout the chain. Therefore, we introduce the concept of reconnaissance in two phases: \emph{external reconnaissance}, which is performed to collect technical or non-technical information before gaining access to an internal asset; and \emph{internal reconnaissance}, which is performed to obtain system information from the internal network.}

\begin{figure}[htb]
    \centering
    \includegraphics[width=\textwidth]{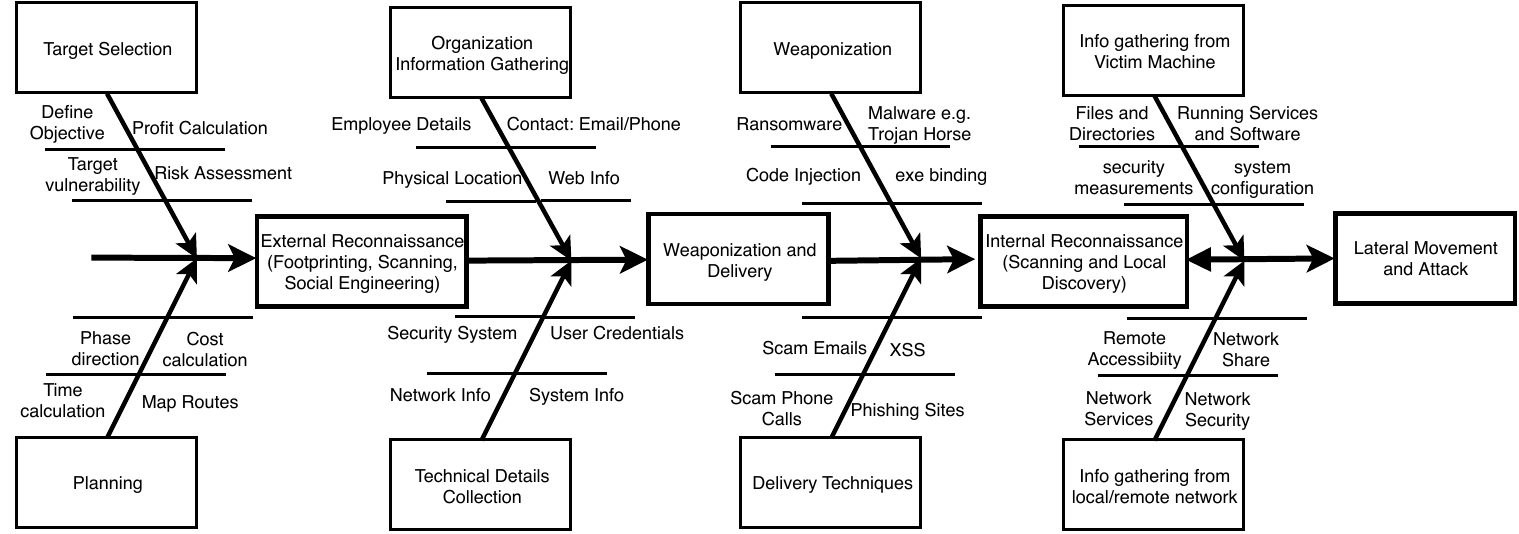}
    \caption{External and Internal Reconnaissance}
    \label{fig:recon_in_cyber_chain}
\end{figure}

Figure~\ref{fig:recon_in_cyber_chain} shows a detailed view of external and internal reconnaissance phases in 
the cyber kill chain model, focusing on large-scale attacks that are carried out by  APTs. 
The attack process starts with target selection and planning. The adversary begins collecting information about the target organization utilizing various footprinting, scanning, and social engineering techniques. Next, the adversary attempts to gain an initial foothold by compromising the target and installing malware or establishing command-and-control through other means.
Then, the adversary can perform internal reconnaissance utilizing various scanning (e.g., active host or port scan) and localhost discovery (e.g., process discovery on host) techniques. 

We could also categorize reconnaissance from the standpoint of the access level required, ranging from 
\emph{outsider}, through \emph{nearsider}, to \emph{insider}. As an outsider, adversaries perform external reconnaissance and can collect publicly available information (e.g., organization or people information) and limited scan results (e.g., public-facing systems, web server version). As a nearsider, adversaries can plant rogue routers to collect network information and to compromise user machines (e.g., employees' portable devices or computers)~\cite{yang2012active}. 
Sometimes adversaries manage to gain physical access to the assets of the target organization (e.g., posing as an employee or a serviceman using social engineering) or compromise and take control of them remotely, which enables adversaries to act as malicious insiders and perform internal reconnaissance~\cite{kandias2011insider}.

\subsection{External Reconnaissance} 
External reconnaissance refers to activities before adversaries gain access to the internal network. Adversaries can obtain crucial information from public-facing nodes, online footprints, and people, which helps them to prioritize objectives and plan attacks. 
\emph{Open-source intelligence} (OSINT) is one of the primary approaches for performing external reconnaissance.
Technical, organizational, and personal weaknesses may be identified by analyzing public sources of information~\cite{tabatabaei2016osint,hassan2019gathering}, while remaining undetected.

%



Adversaries often start by collecting organization information and people's contact details. They can learn different technical details using Internet footprinting, which requires passive techniques with little threat of detection~\cite{mansfield2009simple, hassan2018technical}. However, Internet footprinting tends to provide limited information, so  adversaries may also use social engineering techniques to manipulate people into providing additional information~\cite{albladi2018user, nelson2001methods, krombholz2015advanced}. Attackers next move their attention to designing attacks and malware, and try to compromise at least one internal host. After they have succeeded, adversaries can stay inside the network for months, performing internal reconnaissance and escalating their attacks until they reach their targets~\cite{yadav2015technical}.

\subsection{Internal Reconnaissance}
Once an attacker has compromised at least one host inside the target network or has established insider access, they may create a secure channel between an installed backdoor and a command and control (C2) server~\cite{zeidanloo2009botnet}. The next steps and objectives depend on the information that the adversary can gain from the compromised host. Initially, adversaries can look for user and host-level information. 
Running processes and configurations expose the list of installed software and applications used by the victim host and other hosts~\cite{bartlett2007understanding}. 
They can use system commands and custom tools to collect  user, host, network, and application-level information.
Sometimes, adversaries wait and utilize passive scanning techniques such as sniffing packets to obtain a network view and discover system architectures, protocol mappings, and exploitable vulnerabilities~\cite{hoque2014network}. Passive scanning helps adversaries to remain undetected for extended periods of time. Adversaries can exploit vulnerabilities using the collected information to compromise other hosts to get closer to the target resources~\cite{hoque2014network}. 

\section{Reconnaissance Techniques} \label{sec:recon_techniques}

{We now categorize the most common techniques used for gathering information.}
These techniques are are either \emph{active} or \emph{passive} in nature and used to collect \emph{organization}, \emph{user}, \emph{host}, \emph{network} or \emph{application}-level information. They can be used in either the \emph{external} or \emph{internal} recon~phase.


{Reconnaissance techniques are always evolving, with varying intentions and technical approaches. Generally, reconnaissance can be performed to collect data from different types of sources. For example, information can be obtained from third-party sources at an initial stage, by fooling a target human, or directly from the system resources. In this section, we categorize various reconnaissance techniques and discuss them in the context of the questions that we explored previously: what target information attackers aim to collect (Section~\ref{sec:target_info}) and when they apply the recon techniques  (Section~\ref{sec:phase}).}

\begin{figure}[htb]
    \centering
    \includegraphics[width=.9\textwidth]{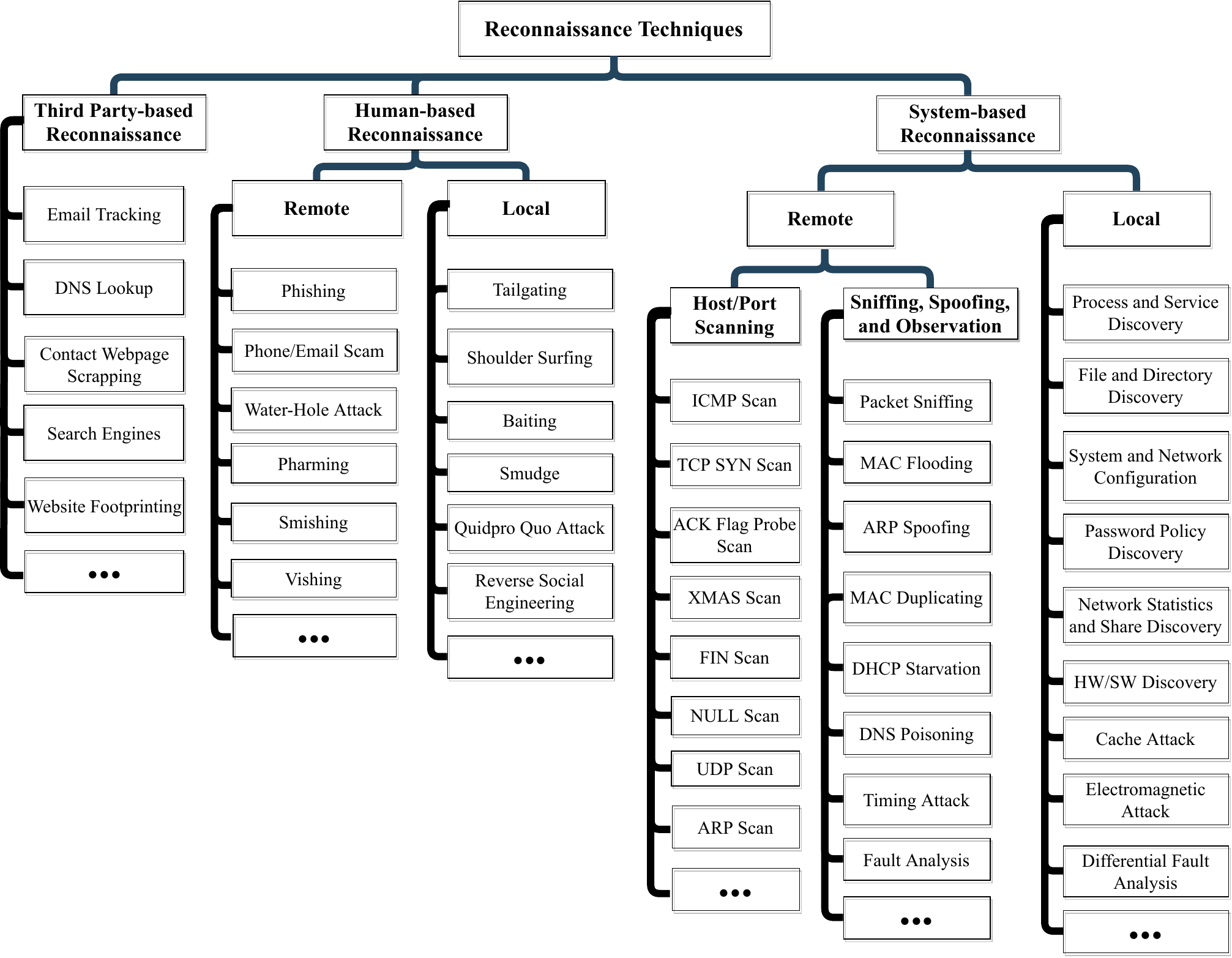}
    \caption{Taxonomy of reconnaissance techniques.}
    \label{fig:recon_techniques}
\end{figure}

{In some cases, adversaries need to interact with the target to obtain information. In other cases, they can obtain information through passive observation or indirect interaction, which is more stealthy. Therefore, many works categorize reconnaissance techniques as either \emph{active} and \emph{passive}. However, while it is possible to categorize some techniques (social engineering, scanning, or side-channels) as either active or passive, there is not always a sharp distinction; so it is not an ideal basis for a comprehensive taxonomy. Instead, we categorize  reconnaissance techniques primarily based on the source of the information: third party-based reconnaissance techniques, human-based reconnaissance techniques, and system-based reconnaissance techniques. Figure~\ref{fig:recon_techniques} lists examples of techniques for each type.}

\begin{itemize}[topsep=2pt, leftmargin=1em]
    \item {\textbf{Third-party source-based reconnaissance techniques:}
    Extracting information from third parties (e.g., third-party websites and services, dark web).}
    \item {\textbf{Human-based reconnaissance techniques:}
    Gathering information from humans by focusing on persons at the target organization.}
    \item {\textbf{System-based reconnaissance techniques:} 
    Collecting information from computer systems (hardware or software) at the target either by exploiting weaknesses or using standard interfaces.}
\end{itemize}


{Third-party source-based and human-based reconnaissance} techniques are usually performed in the external phase, when adversaries look for information about targets prior to launching attacks. 
{System-based reconnaissance} techniques can be applied both externally and internally. {For example,} external scanning gathers information necessary for the initial compromise of the target organization's network. Internal scanning extracts more detailed information regarding the target organization's hosts, networks, services, and applications. However, internal scanning techniques can be riskier for the adversary due to the higher chance of being detected by intrusion detection system~\cite{bou2013cyber}.
Nonetheless, sophisticated APTs can stay hidden inside compromised networks for months to years and perform extensive internal discovery (e.g., Ukraine Power Grid Attack~\cite{case2016analysis}).
In this section, we discuss available tools, outcomes, types of actions (active or passive), and reconnaissance phases (external or internal) for each of the four types.
Table~\ref{tab:passive_footprinting} shows techniques, types, target information, tools, and publicly available tools for {third-party source-based target} footprinting.

\subsection{Third-Party Source-based Reconnaissance}
\label{subsec:footprinting}

{Third-party source-based target} footprinting techniques are typically performed during the early stages of an attack to collect useful information about the organization, personnel, and resources. {Third parties include websites, search engines, dark web, or personnel who are not involved with the target organization.} We discuss the most common {third-party source-based} footprinting techniques here.
{Table~\ref{tab:passive_footprinting} shows techniques, types, target information, tools, and publicly available tools for third-party source-based target footprinting.} 

\begin{itemize}[topsep=2pt, leftmargin=1em]
    \item \textbf{Internet Footprinting:} 
    Adversaries can use tools such as website downloaders, data scrapers, and custom-made scripts to perform Internet footprinting manually. 
    Adversaries often start collecting publicly available technical details
    and then identify underlying technologies~\cite{hassan2018evolution}. For example, an online tool like NetCraft\footnote{\url{https://www.netcraft.com/}}
    is capable of exposing the software and platform behind a website. Site reports contain IP addresses, OS, 
    web server software (e.g., Apache\footnote{\url{https://www.apache.org/}}, IIS\footnote{\url{https://www.iis.net/}}), nameserver, DNS admin, 
    resource specified rules, and site technologies.

    \item \textbf{Whois Lookup:}
    A WHOIS record contains details about the owner of a domain, physical addresses, contact addresses (e.g., telephone numbers and email addresses), and other related information~\cite{wren2010footprinting}.
    WHOIS information is usually stored in WHOIS databases and is maintained by regional Internet registries. 
    The domain registration processes usually require a new domain owner to register with verifiable current contact details. Adversaries can perform WHOIS lookup to find administrative information, including domain name details, the contact information of the owner, name servers, etc. After extracting administrative information, adversaries can perform social engineering attacks to obtain further information about the target. 
    
    \item \textbf{DNS Interrogation:}
    DNS interrogation tools are used to search for hosts in a network {to obtain an interal view of the network}. Several online tools leverage the opportunity to perform a lookup to find additional hosts inside the network. Adversaries can find potential targets by obtaining records of CNAME, PTR, MX, HINFO, and AXFR if misconfigured by administrators~\cite{wren2010footprinting}. NSLookup\footnote{\url{https://linux.die.net/man/1/nslookup}} is the most common tool for DNS interrogation.

    \item \textbf{Website Footprinting:}
    Adversaries can extract typical information such as server and application versions, files, contact details, etc. using website footprinting~\cite{mansfield2009simple}. Footprinting websites is relatively easy since there are many tools available for scanning websites and extracting information (e.g., identifying underlying technology using \snippet{builtwith}\footnote{\url{https://builtwith.com/}}, web crawling using \snippet{HTTrack}\footnote{\url{https://www.httrack.com/}}). Tools like WebExtractor\footnote{\url{http://www.webextractor.com/}}
    can collect contact information, such as phone numbers, email addresses, and fax numbers.
    Other tools, such as Website Watcher\footnote{\url{https://www.aignes.com/}},
    are capable of monitoring web updates. Backdated site information can also be obtained from the Internet~Archive\footnote{\url{https://archive.org/}}.
    
    \item \textbf{Social Media Tracking:}
    Personal information can be obtained through search engines and social media including Facebook, Twitter, and LinkedIn. LinkedIn and other job sites can reveal a person's technical background and responsibility within an organization~\cite{alam2016phishing}.
    Adversaries can follow the online activities of a person and learn about the person's habits, psychological state,  and preferences~\cite{hatfield2019virtuous} for use in social engineering~attacks.
    
    \item \textbf{Email Tracking:}
    Email tracking can include monitoring a user's {time and frequency} of opening and reading emails 
    using publicly available email trackers (e.g., browser extensions such as Streak\footnote{\url{https://www.streak.com/for/email-tracking-in-gmail}}).
    This enables adversaries to learn about their targets' email reading times and associated habits~\cite{englehardt2018never}, which they can exploit in social engineering.
    Initially, they can collect users' email addresses through 
    website footprinting and scraping 
    (e.g., finding contact information on personal websites) or social media scraping (e.g., harvesting user account details from social-media sites). Adversaries can then send malicious links and track if an email was read and if a target followed a link~\cite{krombholz2015advanced}. 
    
    \item \textbf{Search Engines and Google Hacking:}
    Search engines (such as Google, Yahoo, and Bing) can find 
     background information (e.g., financial, technical, or business process reports) about an organization\cite{tabatabaei2016osint}. Google hacking database\footnote{\url{https://www.exploit-db.com/google-hacking-database}} (GHDB) and advanced search queries\footnote{\url{http://www.googleguide.com/advanced_operators_reference.html}} can help adversaries use advanced features of Google search to find more details (e.g., \snippet{filetype} can be used to search specific files).
     In some cases, confidential information including user credentials, vulnerabilities, weaknesses, specific files, etc.\ can be found in GHDB.  Alert services such as Google and Yahoo alerts can track updates of a target website, blog, or media.
\end{itemize}

\begin{table*}[htb]
    \caption{Third-party Footprinting Techniques and Tools}
    \label{tab:passive_footprinting}
    \centering
    \begin{threeparttable}
    \resizebox{.9\columnwidth}{!}{%
    \begin{tabular}{|M{2cm}|M{1.2cm}|M{5cm}|M{1.3cm}|M{5.5cm}|}
        \hline
        Techniques & Type  & Target Information & Phase & Publicly Available Tools\\ \hline \hline
        Internet Footprinting & Passive & Organization Details, People Information  & External & Web tools (e.g. 
        spiderfoot\tnote{1}\space\space),
        search engines (e.g. Google), location (e.g. Google Earth), people (e.g. pipl\tnote{2}\space\space)
        \\ \hline 
        Whois Lookup & Passive  & {User Account Details, DNS and Reachable Hosts} & External & Online tools (e.g. whois \cite{liu2015learning})\\ \hline
        DNS Lookup & Passive  & {DNS and Reachable Hosts, Network View, Network Devices} & External & Online tools 
        (e.g. NSLookup,
        DNSLookUp \cite{hao2010internet})\\ \hline
        Network Footprinting & Active / Passive  & Network  View  & External & Traceroute,
        ARIN DB\tnote{3} \space,
        LoriotPro\tnote{4} \space,
        RIPE\tnote{5} \space,
        LACNIC\tnote{6} \space,
        APNIC\tnote{7} \space,
        and other online tools\\ \hline
        Website Footprinting & Passive  & Organization Details & External & archive.org, website mirroring tools (e.g. NCollector Studio\tnote{8}\space\space
        )\\ \hline
        Email Tracking & Passive  & Contact Details, Account Details & External & Tracking tools (e.g. VisualRoute\tnote{9} \space, GeoSpider\tnote{10}\space\space\space)\\ \hline
        Google Hacking & Passive  & Non-/Technical Details & External & Google advanced search operators \\ \hline
    \end{tabular}}  
    \begin{tablenotes}
    \footnotesize
     \item[1] \url{https://www.spiderfoot.net/}
     $^2$ \url{https://pipl.com/}
     $^3$ \url{https://www.arin.net/resources/guide/account/database/}\\
     $^4$ \url{https://www.loriotpro.com/}
     $^5$ \url{https://www.ripe.net/}
     $^6$ \url{https://www.lacnic.net/}
     $^7$ \url{https://www.apnic.net/}\\
     $^8$ \url{http://www.calluna-software.com/}
     $^9$ \url{http://www.visualroute.com/}
     $^{10}$ \url{http://www.oreware.com/viewprogram.php?prog=22}
   \end{tablenotes}
    \end{threeparttable}
\end{table*}

\subsection{Human-based Reconnaissance}
\label{subsec:social_engineering}

{Human-based reconnaissance, a.k.a.} social engineering (SE), attacks represent some of the most powerful information-gathering techniques according to \emph{Kevin D. Mitnick}~\cite{mitnick2011art}. Typically, fooling a human is significantly easier than fooling firewalls, honeypots, or intrusion detection/prevention system.
Social engineering has been recognized as one of the most common techniques employed by cyber attacks that result in high-profile data breaches (e.g., RSA's SecurID system compromise in 2011 and the New York Times network breach in 2013)~\cite{krombholz2015advanced}.
Social engineering is based on using deception to gain information through {methods like} baiting, pretexting, phishing, and spear-phishing.  We now discuss some of the most common social engineering techniques.

{Existing works categorize social engineering as non-technical vs. technical~\cite{alharthi2020taxonomy} or human-based vs. computer-based~\cite{salahdine2019social}. We categorize based on a similar concept of local vs. remote social engineering techniques. Local SE techniques (e.g., baiting, tailgating, shoulder surfing, etc.) require direct in-person involvement, and remote SE techniques (e.g., phishing, vishing, pharming, malware, etc.) can be performed remotely via web or mobile media.}

\subsubsection{Remote SE Techniques}
{Remote SE techniques are performed remotely using media channels such as mobile, fake websites, spam messages or emails, and malware (e.g., trojan horses or ransomware). These techniques are more common than local SE techniques.}

\begin{itemize}[topsep=2pt, leftmargin=1em]
    \item \textbf{Phishing:}
    Phishing has proven to be a very effective technique for stealing user credentials~\cite{heartfield2015taxonomy}. In a recent paper, Chiew et al.\ presents linkages between media, vectors, and technical approaches of phishing techniques that provide a better understanding of why phishing has been so successful over the years \cite{chiew2018survey}. The authors note the 
    Internet, 
    short messaging service (SMS), eFax, instant messaging, 
    social networking, and telephone services as the primary media of phishing.
    Adversaries can also utilize \emph{evil-twin} attacks, where they lure a target user to connect to a fake wireless access point and authenticate to a forged server so that adversaries obtain the user credentials~\cite{yang2012active}.
    
    \item \textbf{Watering Hole:}
    A watering hole attack {typically compromises} a victim's machine by installing malicious code from a malicious website~\cite{alrwais2016catching}. Adversaries start by profiling a target user or {group} to learn their habits, such as visits to popular websites. Then, the adversaries exploit vulnerabilities in those websites or place links that redirect the users to a malicious site. 
    Since users trust these websites, they may fall victim by accepting downloads or by following malicious hyperlinks allowing attackers to gain access to the victims' machines. 
    Once an adversary has compromised the host (e.g., by installing Trojan horse malware), they can collect user, host, network, and application-level information. 
    
    \item \textbf{Pretexting and Vishing:}
    Pretexting and vishing refer to impersonation through text messages or voice calls (vishing) and convincing targets \ to give access to particular resources~\cite{luo2013investigating,yeboah2014phishing}. 
    For example, an adversary can call a bank pretending to be a trusted person, and convince the official to grant access or to disclose usernames and passwords. Adversaries may require some confidential information to perform this type of attack convincingly~\cite{wang2012smartphone}. 
    
    \item \textbf{Pharming:}
    Pharming is similar to 
    phishing in terms of tempting a target user to visit a fake webpage, but it is more sophisticated technologically since it typically involves secretly installing malicious software on the victim's computer~\cite{brody2007phishing}. Pharming is often performed through DNS poisoning, which enables redirecting victim users to malicious sites even if they attempt to visit only legitimate sites~\cite{stamm2007drive}. Therefore, regardless of the security measures taken by a user they may still fall victim to visiting malicious content.
    
    \item \textbf{Smishing:}
    Smishing (a combination of the words ``SMS'' and ``phishing.'') is a form of phishing in which a victim receives a malicious link in an SMS
    message~\cite{kang2014security}. The victim is tempted to download and install a Trojan horse, keylogger, or some other malware on the victim's mobile phone by following the link in the received message. Several Trojan horses feature keyloggers, which record every keystroke and send the records back to adversaries when the device is connected to the Internet. Account information, credentials, search habits, etc.\ can be obtained from a keylogger-infected machine~\cite{bhardwaj2020keyloggers}.    
\end{itemize}

\subsubsection{Local SE Techniques}
{Local SE techniques involve in-person direct or indirect interaction, such as talking face-to-face, following a person to access a building, or fooling the target by impersonating an authorized person.}

\begin{itemize}[topsep=2pt, leftmargin=1em]
    \item \textbf{Tailgating:}
    {A tailgating attack is effective for attackers to have physical access to an organization or a resource. For example, an attacker can pretend to forget to bring his card and manipulate the target to give him access to a building or secure zone~\cite{salahdine2019social,alharthi2020taxonomy}. RFID card attacks are also common now since many organizations use these as an access token due to low cost and good user experience. However, an attacker can manipulate the RFID network and gain access to the target secure zone~\cite{salahdine2019social}.}
    
    \item \textbf{Shoulder Surfing and Smudge:}
    {An attacker can watch the target person entering a username, passwords, credit information, or other sensitive information by standing near them~\cite{salahdine2019social,alharthi2020taxonomy}. The attacker can also retrieve user input from touch screen devices in the absence of the target person in a \emph{Smudge attack}~\cite{alharthi2020taxonomy}.}

    \item \textbf{Baiting:}
    Baiting is an effective technique for obtaining information by
    spreading Trojan horses using physical media such as flash drives, CD/DVD-ROMs, memory cards, or other portable devices~\cite{krombholz2015advanced}. Usually, the infected media are left in places where target users can find them. {If they} insert the media into their machines due to curiosity or the intention to return the media, {this can result} in infecting the victims' machines and {creating} backdoors for adversaries. Using a keylogger and reverse shell, adversaries can obtain sensitive information from an infected host. In most scenarios, adversaries combine exploits with regular files, so that a victim does not suspect the bait~\cite{stevens2011malicious}.  
    
    \item \textbf{Reverse Social Engineering and Quid pro Quo}:
    Reverse social engineering is another way of manipulating victims to give away confidential information or to let the adversaries gain access. Rick Nelson describes three parts of reverse social engineering: sabotage, advertising, and assisting~\cite{nelson2001methods}.
    Adversaries initiate the process by corrupting or damaging a particular device or workstation. Then, they show advertisements saying that they are capable of fixing it; when the victim asks for help, they extract target information during repair.
    Quid pro quo attacks are a form of reverse social engineering where adversaries call or send messages to random people at the target organization, asking if they requested technical support in the hope that they will eventually contact a person who did~\cite{ivaturi2011taxonomy}.
\end{itemize}

Table~\ref{tab:active_footprinting} shows {approaches}, typical target information,  typical phases, {and types} of common social engineering techniques.
{Approach} refers to whether a particular social engineering technique is active or passive. Social engineering techniques are usually utilized in the external phase and are quite effective in terms of collecting confidential user credentials or other sensitive information: around $85\%$ of  organizations have faced phishing or other social engineering attacks in 2019, which is 
$16\%$ higher than in the previous year~\cite{accentur6:online}.

\begin{table*}[t]
    \caption{Social Engineering Techniques}
    \label{tab:active_footprinting}
    \centering
    \resizebox{\columnwidth}{!}{%
    \begin{tabular}{|M{3cm}|c|M{7.8cm}|M{2cm}|c|}
        \hline
        Techniques & {Approach} & Target Information & Phase & Type\\ \hline \hline
        Phishing, Whaling Attack & Active/ Passive & {User Credentials, Contact/Account Details} & External &  {Remote}\\ \hline
        Watering hole Attack & Active &  {User/Host/Network/Application Information}  & External & {Remote}\\ \hline
        Pretexting and Vishing & Active & {User credentials, Organization Infrastructure, Physical Security}  & External & {Remote}\\ \hline
        Baiting and Quid Pro Quo attacks & Active / Passive & {User/Host/Network/Application Information} & External / Internal & {Local}\\ \hline
        {Tailgating} & {Active} & {Organization Infrastructure, Physical Security} & {External} & {Local}\\ \hline
        Reverse Social Engineering & Active / Passive & User Credentials & External & {Local}\\ \hline
    \end{tabular}}
\end{table*}

\subsection{System-based Reconnaissance}

{System-based recon techniques can be categorized into \emph{remote}  and \emph{local}  information gathering techniques. Adversaries can perform scanning (e.g., TCP, UDP, or ICMP scans) and sniffing (often with the help of, e.g., MAC flooding or ARP spoofing) techniques in a network remotely. Local recon techniques, on the other hand, include discoveries within a compromised host by reading file contents or using operating system commands to explore configurations.}

System-based recon techniques can involve gathering information by directly or indirectly interacting with a system. For example, an attacker can directly scan active hosts by interacting with the hosts (e.g., sending TCP SYN packets to them). The attacker can also gather information indirectly, without interacting with the target (e.g., observing or monitoring leaked information).

\subsubsection{Remote System-based Reconnaissance Techniques}
\label{subsec:scanning}

{Adversaries can perform remote reconnaissance techniques from a remote location to gather information using direct or indirect interaction with a system.}
Network scanning {and sniffing are} 
performed to discover active network resources from an external network or within an internal network. Effective scanning techniques often enable adversaries to find vulnerabilities and to compromise IT assets~\cite{barnett2008towards}.
This information can then be mapped to, e.g., a Common Vulnerability and Exposure (CVE) database, which provides detailed information about publicly known vulnerabilities. Databases and categorization of CVEs are available at MITRE\footnote{\url{https://cve.mitre.org/}},
the National Vulnerability Database\footnote{\url{https://nvd.nist.gov/}},
CVE Details\footnote{\url{https://www.cvedetails.com/}},
etc. {Sniffing techniques are primarily used to capture network packets that reveal sensitive information such as user credentials and protocols being used in the network. One significant distinction between scanning and sniffing is that scanning techniques require direct interaction with the target system, while sniffing uses indirect interaction.}

\paragraph{\textbf{Scanning Techniques}}
{Achleitner et al. categorized malicious network scanning based on the
process of selecting addresses from a scanning space (e.g., IP address space)~\cite{achleitner2016cyber}. According to the authors, 
network scanning includes
uniform scanning (probing random hosts within a IP range), local-preference (preferring a particular region), preference-sequential (probing IP addresses sequentially), non-preference sequential (selecting random IP ranges), and preference-parallel (performing parallel scans). }

Scanning techniques can  be categorized as \emph{stealthy} or \emph{non-stealthy} scanning. 
With stealthy scanning techniques, adversaries
leave minimal trace of the scan and it's origin, which makes stealthy scanning difficult to detect using conventional security measures. Non-stealthy scans are more ``aggressive,'' and there is 
greater chance of being detected by an IDS. 
%
%
%
Stealthy scanning by bots is one of the most sophisticated techniques to efficiently gather information about a network~\cite{dainotti2014analysis}. 
Botnets can be configured to perform a variety of scan types, including uniform scanning where every host is scanned with equal probability~\cite{achleitner2017deceiving}, sequential scanning that systematically explores a space of IP addresses and/or ports~\cite{achleitner2017deceiving}, and preferential scanning which uses additional information to bias the search to specific parts of the network, types of hosts, or ports~\cite{abu2006multifaceted}.
%
%
%
%
Botnet-based stealthy scanning is useful for discovering and compromising network infrastructure while minimizing detection by scanning from many hosts over multiple days~\cite{dainotti2014analysis}.

Scanning techniques can also be categorized as \emph{horizontal scans}, \emph{vertical scans}, and \emph{coordinated/distributed scans}~\cite{barnett2008towards}. If an adversary targets multiple ports on a single IP address, the scan is vertical. A horizontal scan involves targeting a specific port on multiple IP addresses. A coordinated or distributed scan is a combination of both horizontal and vertical scans and can be launched from multiple scanning hosts (e.g., botnet-based scanning). 

First, we discuss some of the most common \emph{{low-level (i.e., network or transport layer)} scanning techniques}, emphasizing the network packet~attributes.


\begin{itemize}[topsep=2pt, leftmargin=1em]
    \item \textbf{TCP Scan with SYN/ACK Flag:} There are several TCP scanning techniques that use SYN or ACK flags to scan a network. TCP SYN scan is a widely used scanning technique; it does not establish a full connection, which makes it relatively stealthy and fast. Adversaries can use the ACK flag to identify open ports as well.

\begin{itemize}[topsep=2pt, leftmargin=1em]
    \item TCP Connect:
    TCP connect scan establishes a full 3-way handshake with hosts within the target IP range~\cite{bhuyan2011surveying}. It starts by sending a SYN packet from a client to the target host. The server responds with a SYN$|$ACK packet (RST packet is sent if the port is closed). Finally, the client sends an ACK in return, establishing the full connection. TCP connect is the simplest scanning technique, and it can be performed without admin privileges since
    it scans active ports, which does not require any special flag settings. However, this scan increases the chance of being detected by an IDS due to establishing an active session~\cite{bhuyan2011surveying}. 
    
    \item TCP SYN Scan:
    SYN scan is a common scanning technique for identifying open and closed ports. SYN scan is also called a \emph{half-open} scanning technique since it does not establish a full TCP connection~\cite{lyon2009nmap}. 
    A SYN scan can be performed quickly within a given range of ports, and it is a relatively stealthy technique. To perform this scan, adversaries send a SYN packet to the target host, and wait to receive the response. If a \emph{SYN} or \emph{ACK} is received, the port is open. If the response is \emph{RST (reset)}, then the port is closed.
    
    \item ACK Flag Probe Scan:
    This scanning technique sets the \emph{ACK} flag instead of the \emph{SYN} flag and determines if a port is open, closed, or unfiltered by analyzing the Time-To-Live (TTL) and window fields within the RST packet header~\cite{NetworkS92:online}. The target port is open if the \emph{TTL} value is less than $64$ or if the window value is not 0. Further, an ACK flag probe may also be able to differentiate between the presence of a stateful or stateless firewall and filtering rules by checking the response or error message (e.g., destination unreachable)~\cite{lyon2009nmap}.
\end{itemize}


\item{\textbf{TCP Scan based on RST Response:}}
Adversaries can set or unset several flags (e.g., FIN, PSH, URG) to perform stealthy scanning. Receiving a packet with \emph{RST} means the port is closed; otherwise, it is open. A popular example of setting the flags is \emph{XMAS} Scan. An inverse TCP scan sets either one flag or none in a TCP packet and is similar to XMAS Scan in terms of detecting open or closed~ports.

\begin{itemize}[topsep=2pt, leftmargin=1em]
    \item XMAS Scan:
    XMAS scan is used to identify ports with the status open and closed \cite{Understa30:online}. The scan involves manipulating the \emph{PSH}, \emph{URG}, and \emph{FIN} flags of a TCP header in crafted packets. 
    An XMAS scan may bypass firewall and ACL filters, and it is fast as well~\cite{lyon2009nmap}. It is called ``XMAS scan'' because if the packet is viewed within \emph{Wireshark}, the enabled alternating bits look like a XMAS-tree.

    \item FIN Scan: FIN scan is also a stealthy scanning technique, similar to the XMAS scan. However, only the \emph{FIN} flag is set~\cite{de1999review}.
    
    \item NULL Scan:
    NULL scan is a stealthy technique similar to XMAS and FIN scanning techniques, but no flag is set in the packet~\cite{de1999review}. The result is the same: ignored packet means open ports, while an \emph{RST} response indicates that the corresponding port is closed. 

\end{itemize}


\item{\textbf{UDP Scan:}}
UDP is simpler than TCP and does not provide the same variety of flag modification schemes as TCP does. However, a UDP scan can still be used to scan open UDP ports that provide a running service. In a UDP scan, a response is typically received if the port is closed.
Typical open services such as DNS, VPN, SNMP, NTP, etc.\ can be determined using UDP port scan~\cite{lyon2009nmap}. In some cases, it is possible to detect versions of services and operating systems as well~\cite{lyon2009nmap}.
{\emph{Listing scanning} is another form of UDP scan that lists IP addresses and names by discovering hosts indirectly~\cite{NetworkS92:online}. The technique involves performing a reverse DNS resolution to determine hostnames.}

\item{\textbf{ICMP Scan:}}
A simple ICMP scan is performed to identify an active network device given a particular 
IP address~\cite{arkin2001icmp}. 
An ``ICMP Covering Ping Sweep'' can 
discover active hosts within a range of IP addresses and can list active nodes based on the subnets~\cite{arkin1999network}.

\item{\textbf{ARP Scan:}}
ARP scanning is a network discovery technique that works by broadcasting an ARP packet in the network and checking which hosts respond~\cite{modi2013survey}. 
Hosts that respond to the broadcast message are active hosts. The ARP scan is a low-level scanning technique that works in local area networks and is usually used to obtain both physical (MAC address) and logical (IPv4/6) addresses of active hosts.
\end{itemize}

{Adversaries may be able to perform {TCP,} {UDP,} and {ICMP} scans from an external network since all of these techniques are routable. Since {ARP} scan is non-routable, adversaries can perform it only in a local area network.
Adversaries can start scanning hosts and ports locally once they have at least one compromised host in the target network.}

Adversaries can also vary the attributes of network scans, including the speed, distribution, and destination of scanning~\cite{barnett2008towards}. Depending on their motivations and on the defenses of the networks, 
adversaries may prefer a \emph{slow scan} approach to avoid detection~\cite{claypool2002stealth}. 
For example, if a port scanner is scanning a host with ports ranging from $1-1024$ and with a time interval of 5 minutes between each port,
performing the scan will take approximately 85 hours.
It is harder for defenders to match and trace these suspicious packets in a vast dataset of traffic over a longer period in a large enterprise system.  

\begin{table}[htb]
    \caption{Network/Transport Layer Scanning Techniques and Tools}
    \label{tab:scanning}
    \centering
    \begin{threeparttable}
    \resizebox{.9\columnwidth}{!}{%
    \begin{tabular}{|M{4cm}|M{1.4cm}|M{3cm}|M{1.4cm}|M{4cm}|}
        \hline
        Techniques & {Approach}  & Target Information & Phase & Tools \\ \hline \hline
        ICMP / TCP / UDP Scanning   & Active  & {Network View/Security} & Internal / External &  NMap\tnote{1} \\ \hline
        Ping Sweep & Active  & {Network View} & Internal / External & NMap, Angry IP scanner\tnote{2}\space, Solarwinds tools\tnote{3}\\ \hline
        ARP Scanning & Active & {Network View} & Internal & ARP Ping\tnote{4} \\\hline
        
        Custom packets using TCP flags (SYN/ACK/FIN scan, XMAS scan, NULL scan) & Active  & {System Services, Network Security} & Internal / External & NMap, Hping2/ Hping3\tnote{6}\space, Amap\tnote{7}\space, SuperScan\tnote{8}\\ \hline
        UDP Scan & Active & {DNS, Network View, System Services} & Internal / External & Nmap\\\hline
        
    \end{tabular}}
    \begin{tablenotes}
    \footnotesize
     \item[1] \url{https://nmap.org/}
     $^2$ \url{https://angryip.org/}
     $^3$ \url{https://www.solarwinds.com/engineers-toolset/use-cases/network-monitoring-tools}\\
     $^4$ \url{https://www.netscantools.com/nstpro_arpping.html}
     $^5$ \url{https://www.ettercap-project.org/}
     $^6$ \url{http://www.hping.org/}\\
     $^7$ \url{https://tools.kali.org/information-gathering/amap}
     $^8$ \url{https://sectools.org/tool/superscan/}
   \end{tablenotes}
    \end{threeparttable}
\end{table}

Table~\ref{tab:scanning} shows the {approach}, target information, phases, and examples of publicly available tools for {scanning} techniques.
{Scanning techniques include ICMP, UDP, ARP, or TCP scanning techniques.} \emph{Type} refers to whether the techniques are active or passive. \emph{Target information} is what adversaries are looking for using these techniques. \emph{Phase} denotes if a particular technique is utilized in external or internal phase. 
Finally, we include publicly available \emph{tools} that are used by security researchers as references. 
However, adversaries may use more sophisticated techniques, such as exploiting services or software vulnerabilities without crashing, performing reconnaissance as regular users, etc. to avoid detection~\cite{stojanovic2020apt}.

{Attackers can also perform \emph{application-level scanning techniques}, such as banner grabbing, operating system and application fingerprinting. Here, we discuss some of the common techniques.}
{Table~\ref{tab:vuln_analysis} presents the {approach}, target information, phases, and examples of publicly available tools for different application-level scanning techniques.}

\begin{itemize}[topsep=2pt, leftmargin=1em]
    \item \textbf{Banner Grabbing:}
    Banner grabbing is a vulnerability scanning techniques that uses application banner information, including name and version~\cite{shamsi2014hershel}.  There are two types of banner grabbing: active and passive. Active banner grabbing requires establishing TCP connections with a remote host to send crafted packets. Adversaries then receive and process the response. Passive banner grabbing involves passive sniffing techniques to capture and analyze network packets. 
    Active banner grabbing techniques are more prone to detection by the defender. Adversaries usually target service ports, such as HTTP, FTP, and SMTP services (ports 80, 21, and 25 respectively). Using banner grabbing techniques adversaries can potentially map an entire
    network~\cite{bajpai2018art}.
    
    \item \textbf{Fingerprinting:}
    Fingerprinting is a method of analyzing response packets to determine the operating system, application version (e.g., web server), or network protocol (e.g., SNMP). 
    Often, the operating system and/or the application reply with packets that expose the platform and version in the packet header.  Adversaries can analyze the response packets, compare the values against a dataset of various operating systems and versions, and identify the OS version (e.g., APT32 \cite{Cybereas65:online}). Information can also be obtained by examining error-message responses. 
\end{itemize}

\begin{table}[htb]
    \caption{Application-level Scanning Techniques and Tools}
    \label{tab:vuln_analysis}
    \centering
    \begin{threeparttable}
    \resizebox{.9\columnwidth}{!}{%
    \begin{tabular}{|M{4cm}|M{1.4cm}|M{3cm}|M{1.4cm}|M{4cm}|}
        \hline
        Techniques & {Approach}  & Target Information & Phase & Tools \\ \hline \hline
        Banner grabbing and OS fingerprinting by sending crafted packets and analyzing responses & Active  & 
        \begin{tabular}[c]{@{}c@{}}{System/Service}\\{Configurations,}\\ {Applications Versions}\end{tabular}
        & Internal / External & Telnet, NetCraft, IDServe\tnote{1} , Nmap, Winfingerprint\tnote{2} \space, Xprobe2\tnote{3} \\
        \hline
        Fingerprinting and patch-level assessment & Active / Passive  & {Host/Network/ Application Vulnerabilities} & Internal / External & Nessus, Saint\tnote{9} \space , Cisco-Torch\tnote{10} \space\space and other vulnerability scanning tools\\ \hline
    \end{tabular}}
    \begin{tablenotes}
    \footnotesize
     \item[1] \url{https://www.grc.com/id/idserve.htm}\\
     $^{2}$ \url{https://securiteam.com/tools/5HP0A1P2LK/}
     $^{3}$ \url{https://github.com/binarytrails/xprobe2}
     $^{4}$ \url{http://lcamtuf.coredump.cx/p0f3/\#}\\
     $^{5}$ \url{https://www.netresec.com/?page=networkminer}\\
     $^{6}$ \url{https://www.securitywizardry.com/products/scanning-products/wireless-tools/netsleuth}\\
     $^{7}$ \url{https://github.com/gamelinux/prads}
     $^{8}$ \url{https://github.com/xnih/satori}\\
     $^{9}$ \url{https://www.saintcorporation.com/products/penetration-testing/}
     $^{10}$ \url{https://tools.kali.org/information-gathering/cisco-torch}
   \end{tablenotes}
    \end{threeparttable}
\end{table}

\paragraph{\textbf{Sniffing Techniques}}

Adversaries can perform sniffing to capture and analyze unencrypted network packets~\cite{cheswick2003firewalls} to collect information like user credentials, e.g., usernames and passwords sent in plaintext. Network packets may also contain information about installed operating systems, applications, protocol versions, source, and destination ports, packet and frame sequences, etc. By analyzing packets frame-by-frame, adversaries may be able to find misconfigurations and vulnerabilities in services. Some protocols are particularly vulnerable to sniffing; for example, Telnet can expose keystrokes (names and passwords), HTTP can reveal data sent in clear texts, SMTP/NMTP/POP/FTP/IMAP can reveal passwords or data sent in cleartext.

Sniffers usually operate in the data link layer of the OSI model. The objective is to compromise the communication channel before the defender in the upper layers is aware and prevents attacks. Attackers often place physical hardware sniffers or network analyzers if they can manage physical access (or a malicious insider) to an organization network (e.g., connect to the SPAN port of a switch that broadcasts all incoming or outgoing traffic).

Passive sniffing or directly capturing packets is performed for discovering network protocols and services, as well as active hosts and ports~\cite{cheswick2003firewalls}. 
Many packet capturing and analysis tools are available on the market; for example, SolarWinds Network Performance Monitor\footnote{\url{https://www.solarwinds.com/network-performance-monitor}}, ManageEngine NetFlow Analyzer\footnote{\url{https://www.manageengine.com/products/netflow/}}, tcpdump\footnote{\url{https://www.tcpdump.org/}}, WinDump\footnote{\url{https://www.winpcap.org/windump/}}, and Wireshark\footnote{\url{https://www.wireshark.org/}}.
These are publicly available tools marketed to network admins, but may be used by adversaries as well. Adversaries can also perform scans using tools and scripts that are customized for a particular vulnerability to remain undetected for a longer period~\cite{ussath2016advanced}.

Active sniffing involves traffic flooding or spoofing attacks to capture traffic or redirect the traffic towards a host controlled by the attacker. Active sniffing is usually performed in a switched network where the attacker might need to use these techniques to capture network traffic.

\begin{itemize}[topsep=2pt, leftmargin=1em]
    \item \change{}{\textbf{MAC Flooding:} MAC flooding involves flooding a switch with abundant mapping requests so that the switch overflows at some point~\cite{ostapenko2013denial}. Eventually, the switch acts as a hub and starts broadcasting all packets, making it easy for the attacker to capture packets.}
    \item \change{}{\textbf{ARP Spoofing:} In this techniques, the attacker usually generates a lot of forged ARP requests and reply packets to flood a switch. When flooded with spoofed ARP requests, the switch is set to ``forwarding mode'' and it is easier for the attacker to capture packets. The attacker can also try to poison the target's ARP table with forged entries that eventually lead to sophisticated attacks like Denial-of-Service and man-in-the-middle (MITM)~\cite{ramachandran2005detecting}. }
    \item \change{}{\textbf{MAC Duplicating/Spoofing:} The attacker can spoof the MAC address of an active target~\cite{anu2017survey}. By duplicating the MAC address, the attacker can take over someone's identity. The technique is useful to gain access to the network if the target MAC address is used to authorize network access. However, this attack is easily detectable by the defender.}
    \item \change{}{\textbf{DHCP Starvation:} In this technique, the attacker sends ``DHCP discovery'' to the routers and attempts to lease all the available IP addresses~\cite{mukhtar2012mitigation}. DHCP starvation is sort of a Denial-of-Service (DoS) attack using DHCP requests. The primary reason for using this technique is to set up a rogue DHCP server that provides IP addresses to others joining the network. Then the attacker can establish the wrong IP, gateway, or DNS servers; used to capture packets.}
    \item \change{}{\textbf{DNS Poisoning:} DNS poisoning is performed by tricking a DNS server into believing the attacker has authentic information that allows the attacker to replace valid IP address entries with fake entries~\cite{anu2017survey}. For example, the attacker can replace a valid IP entry with the IP of a fraud or a phishing site for social engineering or stealing information. The attacker can perform a DNS poisoning attack in two ways: within an internal network, aka intranet (LAN), or replace entries stored in a proxy server. DNS poisoning helps the attacker to bypass security toolbars and phishing filters~\cite{abu2008bypassing}.}
\end{itemize}

Table~\ref{tab:sniffing} presents the approach, target information, phases, and examples of publicly available tools for different sniffing techniques. Passive sniffing refers to listening to the network traffic where the active sniffing techniques are used to enable attacker capture packets in a switched network. Some of these techniques can be performed both externally or internally; other techniques are used within the local area network. Some remote side-channel attacks (e.g., timing or fault analysis) are used to reveal information by sending payloads and then analyzing the responses.

\begin{table}[htb]
    \caption{Sniffing Techniques and Tools}
    \label{tab:sniffing}
    \centering
    \begin{threeparttable}
    \resizebox{.9\columnwidth}{!}{%
    \begin{tabular}{|M{4cm}|M{1.4cm}|M{3cm}|M{1.4cm}|M{4cm}|}
        \hline
        Techniques & \change{Type}{Approach}  & Target Information & Phase & Tools \\ \hline \hline
        MAC Flooding   & Active  & \multirow{9}{2.5cm}{\begin{tabular}[c]{@{}c@{}}Running Protocols\\ and Services,\\ User Data,\\User Credentials\end{tabular}} & Internal &  macof\tnote{1}, yersinia\tnote{2}
        \\ \cline{1-2}\cline{4-5}
        ARP Spoofing & Active  &  & Internal & WinARPAttacker\tnote{4}\space, Cain \& Abel\tnote{3}\space,  UfaSoft Snif\tnote{5} \\ \cline{1-2}\cline{4-5}
        MAC Duplicating & Active &  & Internal & macchanger\tnote{6}\\\cline{1-2}\cline{4-5}
        Network Traffic Sniffing & Passive  &  & Internal / External & Wireshark, Ettercap, TCPdump, Windump
        \\ \cline{1-2}\cline{4-5}        
        DHCP Starvation & Active  &  & Internal & Gobbler\tnote{7}\\ \cline{1-2}\cline{4-5}
        DNS Poisoning & Active &  & Internal / External & Ettercap\\\hline
    \end{tabular}}
    \begin{tablenotes}
    \footnotesize
     \item[1] \url{https://github.com/WhiteWinterWolf/macof.py}
     $^2$ \url{https://github.com/tomac/yersinia}
     $^3$ \url{https://github.com/xchwarze/Cain}\\
     $^4$ \url{http://www.hacker-soft.net/Soft/Soft_2641.htm}
     $^5$ \url{https://ufasoft.com/sniffer/}\\
     $^6$ \url{https://github.com/alobbs/macchanger}
     $^7$ \url{http://gobbler.sourceforge.net/}
   \end{tablenotes}
    \end{threeparttable}
\end{table}

\begin{itemize}
    \item \textbf{Timing Attack:} Leaked timing information from the CPU or memory can be utilized to determine the secret key of a crypto-system or algorithm (e.g., elliptic curve scalar multiplication algorithms). The time samples are gathered using various inputs and placed into a statistical model that predicts the key with a high degree of certainty~\cite{khan2014side,perianin2021end}.
    \item \textbf{Differential Fault Analysis (DFA):} DFA is used primarily for performing cryptanalysis on several cryptographic algorithms (e.g., DES). To compute the amount of leaked information in a practical DFA attack, the attacker must first analyze the distribution of the leaked information and restrict the keyspace. The secret key can be discovered by using appropriate information estimate modeling~\cite{piret2003differential}.
\end{itemize}



\subsubsection{Local System-based Reconnaissance}
\label{subsec:discovery}

Once adversaries have compromised at least one asset in the target organization, they can start collecting local system information. For example, they can install rootkits, Trojan horses, or other malware that connect back to the command and control servers established by adversaries beforehand~\cite{daly2009advanced}. Adversaries can then remotely execute commands or 
use additional exploits. 

\begin{itemize}[topsep=2pt, leftmargin=1em]
    \item \textbf{User and Group Discovery:} 
    Adversaries can look for system and domain account information to learn about user and group credentials, which they may then use for privilege escalation. On the Windows platform,  commands such as \snippet{net user}, \snippet{net group}, and \snippet{net localgroup} can be used for querying user or group information (e.g., APT1 \cite{APT1Expo92:online}). On  Unix-based systems, \snippet{/etc/passwd} and \snippet{/etc/groups} files are available for querying user and group information. 

    \item \textbf{Process Discovery:}
    On most platforms there are several built-in command tools that can discover running processes on a system. For example, on the Windows platform, a built-in tool named \snippet{tasklist} is available for performing process and security system queries  (e.g., APT: 
    navRAT \cite{TalosBlo98:online}). 
    On Unix-based systems, 
    the built-in command \snippet{ps} is available for checking running processes (e.g., APT:
    XAgentOSX \cite{XAgentOS63:online}).
    
    \item \textbf{Service Discovery:}
    Adversaries can collect information about running services on the Windows platform using system commands like \snippet{net start} (e.g., APT: Sykipot \cite{AnotherS94:online}), \snippet{tasklist} (e.g., APT: Kwampirs \cite{NewOrang81:online}), or \snippet{sc query} (e.g., APT: OilRig \cite{TheOilRi58:online}). On 
    Unix-based systems, they can run system commands like \snippet{service}, \snippet{chkconfig}, or \snippet{netstat} to 
    obtain service-oriented information.

    \item \textbf{Network Configuration Discovery:}    
    Adversaries can look for basic network configuration information such as IP and MAC addresses, network adapters or interface, etc.\ using the commands  \snippet{ipconfig} (e.g., APT: BabyShark \cite{NewBabyS48:online})
    and \snippet{ifconfig} (e.g., APT: Calisto \cite{CalistoT34:online})
    on Windows and Unix-based systems. They can then look for more details including the default gateway, primary and secondary WINS, DHCP configuration, and DNS server details. A number of APTs use \snippet{nbtstat} (e.g., APT: 
    Epic \cite{TheEpicT37:online}) or \snippet{nbtscan} (e.g., APT: Soft Cell \cite{Operatio62:online}) to query NetBIOS name resolution information and to find vulnerabilities (e.g., APT: Turla \cite{Waterbug56:online}). ARP information can be obtained using the command \snippet{arp -a} (e.g., APT: Kwampirs \cite{NewOrang81:online}). Some APTs can
    perform query and enumeration over the ARP cache or table (e.g., APT: Olympic Destroyer~\cite{TalosBlo94:online}).

    \item \textbf{File and Directory Discovery:}
    Adversaries can list directory items on a Windows-based system by running \snippet{dir} or \snippet{tree} command (e.g., APT: 
    BabyShark \cite{NewBabyS48:online}). Adversaries have been reported
    to go through both system configuration files and user-created files~\cite{TheEpicT37:online}. On Unix systems, configuration files can typically be accessed from the \snippet{/etc} directory. Basic commands like \snippet{ls}, \snippet{find}, \snippet{locate} etc.\ are available to search and explore files on Unix systems. On Windows, software information is available in the \snippet{Program Files} directory. Adversaries can use custom scripts that can search for specific files with particular extensions (e.g., APT: 
    Micropsia~\cite{Micropsi75:online}). 

    \item \textbf{Password Policy Discovery:}
    Adversaries can also learn information about the password policies enforced on a system. This is helpful for planning brute-forcing attacks or designing custom password dictionaries. Details such as user password age, password type, or hints can be obtained using user commands, e.g., \snippet{chage -l \$USER} on  Unix or Linux platforms. For the Windows platform, \snippet{net accounts} command provides account password policies (e.g., APT: OilRig \cite{Targeted73:online}); while for macOS, user command \snippet{pwpolicy getaccountpolicies} can be used. On Linux systems, the policies are available in the
    \snippet{/etc/pam.d/common-password} file.

    \item \textbf{Network Statistics Discovery:} 
    If adversaries intend to perform detailed internal scanning later they may initially want to learn network statistics, e.g., local TCP and UDP connections, routing tables, lists of network interfaces, etc.\ using the command line tools \snippet{netstat} (e.g., APT: 
    BlackEnergy \cite{BE2Custo43:online, BlackEne6:online}), \snippet{net use} (e.g., APT: APT1 \cite{APT1Expo92:online}), and \snippet{net session} (e.g., APT: Epic \cite{TheEpicT37:online}). \snippet{netstat -aon} is a common command to gather network connection information; it reveals network connections and can search a specific IP range in a network.

    \item \textbf{Network Share Discovery:}
    Shared directories and files across the network provide access may also contain valuable information. Some APTs are also able to perform enumeration of network shares~\cite{TR12Anal74:online}, which results in gathering potential attack vectors for other systems. \snippet{net view} or \snippet{net share} is used to collect SMB information across Windows platform based networks (e.g., APT: 
    APT41 \cite{ReportDo23:online}). Linux supports both NFS and SMB. \snippet{smbclient}, \snippet{nfsstat -m}, and \snippet{df -aH} commands can be used to explore if a network share is available on the compromised machine.
    
    \item \textbf{Keylogging and Screen Capture:}
    Adversaries can use keyloggers to collect users' key-strokes and information, such as passwords, habits, or financial information~\cite{bhardwaj2020keyloggers}. For example, terminal commands or application names typed in by a user can reveal further details of a system, used applications, and services. Keylogging helps adversaries to monitor host activities passively. 
    Several keyloggers are also capable of recording desktop screens~\cite{bhardwaj2020keyloggers}.
\end{itemize}

Adversaries utilize local {system-based (host)} {reconnaissance} techniques to determine installed software, applications, packages, and frameworks. Configurations and environment variables are relatively easy to discover in a compromised host. Files and directories may contain important and confidential information for further compromise or exhilaration. Internal host discovery can directly or indirectly lead to further exploitation, lateral movement, escalation, or data that is the ultimate target of the attacker.
Table~\ref{tab:local_discovery} presents examples of local discovery techniques, {approach}, target information, phases,
and command-line tools with examples of APTs that use these tools.

\begin{table*}[t]
    \caption{Local System-based (Host) Reconnaissance Techniques}
    \label{tab:local_discovery}
    \centering
    \resizebox{\columnwidth}{!}{%
    \begin{tabular}{|M{2.7cm}|c|M{5.5cm}|c|M{5cm}|M{3.7cm}|}
        \hline
        Techniques & {Approach} & Target Information & Phase & Commands-line Tools & Example APTs\\ \hline \hline
        User and Group Discovery & Active & Account Information  & Internal & \emph{Windows:} \snippet{net user}, \snippet{net group}, \snippet{net localgroup} & Ke3chang~\cite{Exposing59:online}, OilRig~\cite{TheOilRi58:online}\\\hline
        
        Process Discovery & Active & Process  & Internal & \emph{Windows:} \snippet{tasklist}; \emph{Unix:} \snippet{ps} & APT1~\cite{APT1Expo92:online}, BabyShark~\cite{NewBabyS48:online}, ZxShell~\cite{ThreatSp18:online}\\\hline
        
        Service Discovery & Active & System Services, Service Configuration & Internal & \emph{Windows:} \snippet{net start}, \snippet{tasklist}, \snippet{sc query}; \emph{Unix:} \snippet{service}, \snippet{chkconfig}, \snippet{netstat} & APT1~\cite{APT1Expo92:online}, Epic~\cite{TheEpicT37:online}, Ke3chang~\cite{Exposing59:online}, OilRig~\cite{TheOilRi58:online}, Turla~\cite{Waterbug56:online}\\\hline
        
        Network Configuration Discovery & Active & Network View, Peripheral Devices & Internal & \emph{Windows:} \snippet{ipconfig}, \snippet{nbtstat}, \snippet{nbtscan}; \emph{Unix:} \snippet{ifconfig} & APT1~\cite{APT1Expo92:online}, APT32~\cite{Cybereas65:online}, Epic~\cite{TheEpicT37:online}, Turla~\cite{Waterbug56:online}\\\hline
        
        File and Directory Discovery & Active & Files and Directories & Internal & \emph{Windows:} \snippet{dir}, \snippet{tree}; \emph{Unix:} \snippet{ls}, \snippet{find}, \snippet{locate} & admin@338, BabyShark~\cite{NewBabyS48:online}, Elise\\\hline
        
        Password Policy Discovery & Active & System Configuration (Password Policy) & Internal & \emph{Windows:} \snippet{net accounts}; \emph{Unix:} \snippet{chage -l \$USER}; \emph{macOS:} \snippet{pwpolicy getaccountpolicies} & Kwampirs~\cite{NewOrang81:online}, OilRig~\cite{TheOilRi58:online}\\\hline
        
        Network Statistics Discovery & Active & Network Traffics, Host Peripheral Devices & Internal & \emph{Windows:} \snippet{netstat}, \snippet{net use}, \snippet{net session}; \emph{Unix:} \snippet{netstat} & APT1~\cite{APT1Expo92:online}, APT32~\cite{Cybereas65:online}, APT41~\cite{ReportDo23:online}, Epic~\cite{TheEpicT37:online}, Oilrig~\cite{TheOilRi58:online}, Turla~\cite{Waterbug56:online}\\\hline
        
        Network Share Discovery & Active & Shared Files and Directories & Internal & \emph{Windows:} \snippet{net view}, \snippet{net share}; \emph{Unix:} \snippet{smbclient}, \snippet{nfsstat}, \snippet{df} & APT41~\cite{ReportDo23:online}, Kwampirs~\cite{NewOrang81:online}\\\hline
        
        Keylogging and Screen Capturing & Passive & Host I/O Interfacing & Internal & N/A & APT41~\cite{ReportDo23:online}, Oilrig~\cite{TheOilRi58:online}, Turla~\cite{Waterbug56:online}\\\hline
    \end{tabular}}
\end{table*}

{There are several other local system-based reconnaissance techniques that require the attacker to have physical access to the system to observe characteristics of the system. For example, numerous side-channel attacks including power~\cite{prouff2009statistical}, electromagnetic
(EM) emanation~\cite{khan2014side}, and acoustical~\cite{genkin2017acoustic,genkin2014rsa} analyses require physical access to the devices.}

\begin{itemize}
    \item {\textbf{Cache Attack:} A cache attack can be triggered by eavesdropping on keyboard timings, for example, an attack in an address in the GTK Library while processing keystrokes~\cite{gruss2016flush+}. The attack is executed as a program that flushes the address and identifies when a keystroke occurred based on memory access times or the clflush instruction's execution time.   
    Two major CPU vulnerabilities (Meltdown~\cite{lipp2018meltdown} and Spectre~\cite{kocher2019spectre}) can be used to perform cache-based side-channel attacks by leaking sensitive information from the memory.}
    \item {\textbf{Electromagnetic Attack:} An electromagnetic {signal} carries information such as power, time, and so on. Leakage of this information can aid attackers to break into the security system to find out secret keys~\cite{khan2014side}. The leakage of compromising information via electromagnetic (EM) emanations from CMOS devices can lead to attacks on cryptographic devices where the power side-channel is unavailable. Signal Detection and Estimation Theory techniques can be used to combine leakages from several EM channels, resulting in powerful attacks \cite{agrawal2002side}.}
    \item {\textbf{Acoustic Cryptanalysis:}
    Eavesdropping on acoustic emanations can be used to listen in on slow electromechanical components like keyboards and printers to reveal additional data for side channel attacks. 
     Compromised mobile device eavesdropping, eavesdropping bugs, and auditory eavesdropping can all be used to carry out similar attacks~\cite{genkin2017acoustic}. For example, this data can be used to extract information about the CPU operations of laptop computers. An attacker can even discover the commands that the target computer executes by eavesdropping on acoustic emanations with a microphone~\cite{genkin2014rsa}.}
     \item {\textbf{Additional Methods:}
     In addition to the ones listed above there are many other local side-channel attacks such as NAND mirroring, clock or power glitches, temperature variation, smudges, differential computation analysis, etc~\cite{spreitzer2017systematic}. Nearly any feature of a local system is potentially a useful source of side-channel information.}
    
   
\end{itemize}

\section{Conclusion and Future Work}
\label{sec:concl}

Gaining a clearer understanding of how and why cyber adversaries conduct reconnaissance activities is a critical area of research for cyber defense, since successful attacks depend so heavily on effective reconnaissance. However, we find that there is little comprehensive research on this topic to establish a big picture view of how reconnaissance works, including the large variety of methodologies and tools used to conduct reconnaissance. Our first research goal was to establish a broad picture of what types of information adversaries seek.
{Next, we consider when attackers conduct reconnaissance in the standard kill chain model, as well as from what perspective (internal vs. external).}
Our third goal focuses on understanding the {wide} variety of specific techniques and tools adversaries use to gather this information. We  developed taxonomies for both information types and the techniques to help organize these into useful categories. {While we draw inspiration from distinctions previously drawn in the literature, in some cases we find that in some cases these distinctions were too vague or limited to be used for a general taxonomy, so we adopted new dimensions that are clearer and more useful to practitioners.} 

One of the main lessons from our survey is the overall scope and diversity of  adversarial reconnaissance in cybersecurity. The variety of types of information that could potentially be useful to an attacker is vast, as is the number of tools and specific techniques for obtaining it.  This is also a moving target, {since} the types of information that are relevant and the tools will naturally evolve over time with technology.
Nevertheless, {we were able to capture some common distinctions in our taxonomy as well as the analysis and clustering of more specific techniques.} 
{For example, it is important to think broadly about the types of the information that is being collected and the different places it is collected from, including what is publicly accessible and what is not. The techniques used are quite different depending on the source of the information (including the key distinction between technological and social methods), and therefore the relevant defenses are also quite different.}
{Some common features of the techniques such as the spectrum from active to passive are also very important, since these have a direct correlation with the likelihood of detection and when in the attack cycle they are most likely to be deployed.} 
We also observe that the type and objectives of the adversary may have a great impact on how they conduct reconnaissance activities. {We now go into greater detail on some of the specific observations from our study that can lead to areas for improvement and future research in adversarial reconnaissance as well as potential counter-measures.}

\subsection{Improving Adversarial Reconnaissance Models}

{Our survey provides a comprehensive overview of reconnaissance activities; however, we still have a limited understanding of how to model the details of the reconnaissance process of different types of attackers. This includes how they make decisions about what types of reconnaissance to conduct, how to prioritize different types of information, and how they form detailed beliefs about systems and defenses based on limited and uncertain information. There is also a limited understanding of how attackers make key tradeoffs such as utilizing stealthy vs. non-stealthy methods.}

While there are many case studies and examples, we lack a general framework and data to model typical reconnaissance activities. {The formal models that do exist to date (e.g., in the literature on cyber deception) are typically quite simplistic and/or limited in scope (e.g., specifying just one type of scanning procedure and assuming attackers collect perfect information.}
There are some recent studies that have considered evaluating the efficiency of deception in the reconnaissance phase, including models that incorporate Bayesian updating of beliefs~\cite{sugrim2018measuring}. {Another study developed a model of the reconnaissance capabilities of persistent, stealthy adversaries and demonstrated that these adversaries are capable of conducting effective network reconnaissance passively; this offers a method for defining cost and reward criteria that adversaries use to determine which targets to pursue when moving laterally across the network ~\cite{pham2020quantitative}.} Another line of work considers modeling adversary knowledge using a set of logic formulas with probabilities~\cite{jajodia2017probabilistic}.

\subsection{Empirical Studies of Reconnaissance}

{A related issue is the general need for more empirical research to answer basic questions including what the most common types of reconnaissance activities are, how these activities vary across different types of attackers, how attackers make decisions about conducting reconnaissance, how they use this information in attack planning. Such studies are naturally difficult to conduct since attackers actively try to hide much of this information, but there is still a notable lack of good, high-quality data and empirical work to study both the prevalence and effectiveness of different types of reconnaissance approaches in  both controlled and uncontrolled environments. There is also a lack of good metrics and empirical evidence regarding the effectiveness of different defensive mitigation strategies at limited or obfuscating the information attackers can gather, as well as the ability to detect different types of reconnaissance activities. More effective models of the process and beliefs of adversaries will help to scope this type of evaluation, but we also need better sources of real world data and experimental designs to understand how attackers gather information in the real world. 
}


\subsection{Reconnaissance Countermeasures}

Another useful outcome of our survey is to contribute to developing countermeasures that can hinder the ability of adversaries to obtain key information that would enable successful attacks. Network and host-based intrusion detection systems typically monitor for scanning and other known/obvious adversarial reconnaissance activities. 
Many techniques have been proposed in the literature to use deception and information hiding to mitigate reconnaissance, including honeypots, honey tokens, honey passwords, honey permissions and parameters, etc.~\cite{juels2013honeywords,kaghazgaran2015toward,wang2017honey}.
Moving Target Defense (MTD) can also increase the  complexity, diversity, and randomness of the cyber systems for an attacker doing reconnaissance~\cite{wang2018cyber}. Techniques such as dynamic host address translation, route alteration, and IP randomization can lower the success of passive reconnaissance~\cite{achleitner2017deceiving,achleitner2016cyber}. Additional methods including database decoys, OS obfuscation, source code decoys, forging fake traffic, topology deception, hyperlinks decoys, simulation deception, and code embedding deception~\cite{sun2016desir,rowe2004model,salem2011decoy} can mitigate both passive and active reconnaissance. 
{Employee training, security awareness and best practices can mitigate social engineering tactics to some extent. Table~\ref{tab:defensive_measurements}
presents an overview of which types of defensive measures can counter particular reconnaissance techniques.}

\begin{table}[!ht]
\centering
\caption{Defensive Measures against Reconnaissance Techniques}
\label{tab:defensive_measurements}
\resizebox{0.85\textwidth}{!}{%
\begin{tabular}{|c|c|c|c|}
\hline
\diagbox[width=15em]{Measures}{Techniques}     & \begin{tabular}[c]{@{}c@{}}Third-party source-based\end{tabular} & \begin{tabular}[c]{@{}c@{}}Human-based\end{tabular} & \begin{tabular}[c]{@{}c@{}}System-based\end{tabular}\\ \hline\hline
Reconnaissance Detection & \textbf{\textsf{X}} & \checkmark & \checkmark  \\\hline
Cyber Deception/MTD   &  \textbf{\textsf{X}}   &  \textbf{\textsf{X}} &  \checkmark  \\ \hline
Security Awareness and Best Practices & \checkmark  & \checkmark &  \checkmark  \\ \hline
\end{tabular}}
\end{table}

{Future work could elaborate on reconnaissance countermeasures based on each category in the taxonomy. Defenders can detect some techniques more easily; for example, scanning and sniffing in an internal network can be detected using an intrusion detection system. There is a need to evaluate the effectiveness of different strategies to detect the different types of recon techniques. Some techniques (e.g., side-channel attacks) are challenging to detect, and other techniques (e.g., third-party source-based recon) cannot be identified. Therefore, other types of mitigation strategies are necessary, and must also be evaluated. Better models (as discussed above) can help to formalize many of these questions and provide useful evaluation measures, and out taxonomy can help to ensure comprehensive coverage and identify areas with limited mitigation options.}

\subsection{Evolving Forms of Reconnaissance Techniques}

{As technology changes, the nature of reconnaissance also changes due to new types of information becoming relevant as well as new techniques being developed to extract useful information. While we have focused mostly on common current techniques, we note some evolving trends that will affect adversary reconnaissance in the future. One is the rise of disruptive technologies such as virtualization, cloud, fog, and mobile or edge computing, and containerization. One of the effects of this is that organizations often do not control all of their own computing resources and data locally, but outsource them to other vendors who operate cloud resources. This presents new opportunities for social engineering attacks, as well as new types of side channel attacks, for example, cross-VM cache-based~\cite{anwar2017cross}, GPU-based~\cite{naghibijouybari2018rendered}, or directory-based~\cite{yan2019attack}, etc. that have only recently begun to be recognized and considered in the literature~\cite{lyu2018survey}. The increasing complexity for organizations that must operate or source software and hardware across national boundaries and different regulatory jurisdictions is also an issue that will need further consideration.}

{Another rapidly evolving area is the use of artificial intelligence and machine learning methods both in business processes as well as network management and cyber defense. These autonomous agents present a new target for attackers. 
One example of this is accessing valuable data, such as by stealing machine learning models that have been trained on highly valuable data sets~\cite{tramer2016stealing,melicher2016fast,miao2021machine}.  
Another problems is the potential for attackers to fool AI system (including authentication and intrusion detection systems) into making erroneous decisions and providing additional vulnerabilities for system access \cite{pitropakis2019taxonomy}. The means for attackers to learn about these automated systems are only just starting to be explored, and while impressive proofs of concept of the vulnerabilities of these systems have been demonstrated there is limited work done so far to understand the extent, prevalence, and exploitability of these systems in the real world, or to address how attackers can systematically gather reconnaissance information about AI systems to use in~attacks.}

\section*{Acknowledgements}
This material is based upon work supported by the National Science Foundation under Grant CNS-1850510 and by the Army Research Office under Award W911NF-17-1-0370. Any opinions, findings, and conclusions or recommendations expressed in this material are those of the authors and do not necessarily reflect the views of the National Science Foundation or the Army Research~Office.

\bibliographystyle{ACM-Reference-Format}
\bibliography{main}


\end{document}